\PassOptionsToPackage{table}{xcolor} % load colortbl when xcolor is loaded - for coloring tables
\documentclass[%
 reprint,
 amsmath,amssymb,
 aps,
 longbibliography,
floatfix
]{revtex4-1}

\usepackage[tight]{subfigure} % subfigures, revtex compatible
\usepackage{dcolumn}% Align table columns on decimal point
\usepackage{bm}% bold math
\usepackage{color} % colored text
\usepackage[pdftex,colorlinks=true,
allcolors=blue
]{hyperref}% add hypertext capabilities
\usepackage{hypcap}
\usepackage{paralist}% for compactitem used by Marko
\usepackage{multirow}% multicolumn/row cells in tables

\usepackage{url} % add auto-linked urls
\usepackage{mathtools} % shortintertext, and other useful math-related features
\usepackage{siunitx}
\usepackage[dvipsnames]{xcolor}

\usepackage{paralist} % needed for compactenum, compactitem, and inparaenum
\usepackage{soul}

\usepackage[flushleft]{threeparttable} % needed to add notes
\usepackage[table]{xcolor}
\usepackage{array,ragged2e}

\newcommand{\logLogSlopeTriangle}[5]
{
    \pgfplotsextra
    {
        \pgfkeysgetvalue{/pgfplots/xmin}{\xmin}
        \pgfkeysgetvalue{/pgfplots/xmax}{\xmax}
        \pgfkeysgetvalue{/pgfplots/ymin}{\ymin}
        \pgfkeysgetvalue{/pgfplots/ymax}{\ymax}

        \pgfmathsetmacro{\xArel}{#1}
        \pgfmathsetmacro{\yArel}{#3}
        \pgfmathsetmacro{\xBrel}{#1-#2}
        \pgfmathsetmacro{\yBrel}{\yArel}
        \pgfmathsetmacro{\xCrel}{\xArel}

        \pgfmathsetmacro{\lnxB}{\xmin*(1-(#1-#2))+\xmax*(#1-#2)} % in [xmin,xmax].
        \pgfmathsetmacro{\lnxA}{\xmin*(1-#1)+\xmax*#1} % in [xmin,xmax].
        \pgfmathsetmacro{\lnyA}{\ymin*(1-#3)+\ymax*#3} % in [ymin,ymax].
        \pgfmathsetmacro{\lnyC}{\lnyA+#4*(\lnxA-\lnxB)}
        \pgfmathsetmacro{\yCrel}{\lnyC-\ymin)/(\ymax-\ymin)} % THE IMPROVED EXPRESSION WITHOUT 'DIMENSION TOO LARGE' ERROR.

        \coordinate (A) at (rel axis cs:\xArel,\yArel);
        \coordinate (B) at (rel axis cs:\xBrel,\yBrel);
        \coordinate (C) at (rel axis cs:\xCrel,\yCrel);

        \draw[#5]   (A)-- node[pos=0.5,anchor=north] {1}
                    (B)--
                    (C)-- node[pos=0.5,anchor=west] {#4}
                    cycle;
    }
}
\usepackage{graphicx}% Include figure files
\graphicspath{{figs/},{figs_tikz/}}

\usepackage{ifthen}

\def\UseTikz{true}
\def\TikzExternalize{false}

\ifthenelse{\equal{\UseTikz}{true}}{
  \usepackage{pgfplots}
  \usepackage{pgfplotstable}
  \pgfplotsset{compat=newest}
  \usepgfplotslibrary{patchplots}
  \usepgfplotslibrary{colorbrewer} % nice colorschemes
  \usetikzlibrary{colorbrewer}
  \usetikzlibrary{shadows}
  \usetikzlibrary{external}
  \newcommand{\ftbedataset}{figs/ftbe/}
  \pgfplotstableset{search path={\ftbedataset}}
  \usepackage{braids}
}{
  \usepackage{tikzexternal}
}
\ifthenelse{\equal{\TikzExternalize}{true}}{
  \tikzexternalize[prefix=./figs_tikz/]
}{}

\definecolor{sRED}{RGB}{228, 26, 28} % redish
\definecolor{sBLUE}{RGB}{55, 126, 184} % blue-ish
\definecolor{sGREEN}{RGB}{77, 175, 74} % greenish
\definecolor{sPURPLE}{RGB}{152, 78, 163} % purplish
\definecolor{sORANGE}{RGB}{255, 127, 0} % yellowish

\DeclareMathOperator{\ftbe}{FTBE}
\newcommand{\mftbe}[1][n]{\max\ \ftbe}

\DeclarePairedDelimiter{\abs}{\lvert}{\rvert}

\newcommand{\braidlab}{\texttt{braidlab}{}}

\newcommand{\nofrac}[2]{#1/#2} % sometimes nicer to use / fraction

\newcommand{\taudim}{\tilde\tau}
\renewcommand{\Re}{\operatorname{Re}}

\begin{document}

\preprint{APS/123-QED}

\title{Using braids to quantify interface growth and coherence in a rotor-oscillator flow}

%\thanks{A footnote to the article title}%

\author{Margaux Filippi}
\email{margaux@mit.edu}
 \affiliation{Department of Mechanical Engineering, Massachusetts Institute of Technology, Cambridge, MA 02139, U.S.A.}
 \affiliation{Woods Hole Oceanographic Institution, Woods Hole, MA 02543, U.S.A.}

\author{Marko Budi\v{s}i\'{c}}%
%\email{marko@clarkson.edu}
\affiliation{%
 Department of Mathematics, Clarkson University, Potsdam, NY 13699, U.S.A. }

%\collaboration{MUSO Collaboration}%\noaffiliation

\author{Michael R. Allshouse}
 \affiliation{Department of Mechanical and Industrial Engineering, Northeastern University, Boston, MA 02115, U.S.A.}

\author{S\'{e}verine Atis}
 \affiliation{Department of Physics, University of Chicago, Chicago, IL 60637, U.S.A.}

\author{Jean-Luc Thiffeault}
 \affiliation{Department of Mathematics, University of Wisconsin -- Madison, Madison, WI 53706, U.S.A.}

\author{Thomas Peacock}
 \affiliation{Department of Mechanical Engineering, Massachusetts Institute of Technology, Cambridge, MA 02139, U.S.A.}

%\shorttitle{Braids to quantify growth and coherence in a rotor-oscillator flow} %for header on odd pages
%\shortauthor{M. Filippi et al} %for header on even pages

%\corresp{\email{margaux@mit.edu}},

%\homepage{http://www.Second.institution.edu/~Charlie.Author}

%\collaboration{CLEO Collaboration}%\noaffiliation

\date{\today}% It is always \today, today,
             %  but any date may be explicitly specified

\begin{abstract}
The growth rate of material interfaces is an important proxy for mixing and reaction rates in fluid dynamics, and can also be used to identify regions of coherence. Estimating such growth rates can be difficult, since they depend on detailed properties of the velocity field, such as its derivatives, that are hard to measure directly.  When an experiment gives only sparse trajectory data, it is natural to encode planar trajectories as mathematical braids, which are topological objects that contain information on the mixing characteristics of the flow, in particular through their action on topological loops.  We test such braid methods on an experimental system, the rotor-oscillator flow, which is well-described by a theoretical model.  We conduct a series of laboratory experiments to collect particle tracking and particle image velocimetry data, and use the particle tracks to identify regions of coherence within the flow that match the results obtained from the model velocity field.  We then use the data to estimate growth rates of material interface, using both the braid approach and numerical simulations.  The interface growth rates follow similar qualitative trends in both the experiment and model, but have significant quantitative differences, suggesting that the two are not as similar as first seems.  Our results shows that there are challenges in using the braid approach to analyze data, in particular the need for long trajectories, but that these are not insurmountable.
\end{abstract}

%\pacs{Valid PACS appear here}% PACS, the Physics and Astronomy
                             % Classification Scheme.
%\keywords{Suggested keywords}%Use showkeys class option if keyword
                              %display desired
\maketitle

%\tableofcontents

\section{\label{sec:Intro}Introduction}

Characterization of material mixing in fluid flows is an active research area at the intersection of mathematics, engineering and physics. Two common types of analysis are to (1) quantify the overall amount of stirring and mixing that occurs in the system and (2) to identify structures that enhance or reduce material mixing. The earliest examples of a systematic mathematical approach to these problems uses techniques from nonlinear dynamics that rely on having the fluid velocity field as a continuous, often differentiable, function of space and time~\cite{Ottino1989,Aref1984,MacKay1984,Rom-Kedar1990a}.  (See for example~\cite{Sturman2006,Ottino1989,Childress1995,Meiss2015,Balasuriya2016b,Bollt2013,Leonard1987,Aref2017Rev.Mod.Phys.} for comprehensive reviews.)  By now, these techniques have been sufficiently adapted and refined to be applicable to data from real fluid flows that, for example, arise in and around living organisms~\cite{Green2010,Shadden2006,Katija2009,Tan2018}, govern ecological processes~\cite{Thiffeault2010a,Mezic2010,Olascoaga2012} and arise in engineered systems and technological processes~\cite{Weldon2008,Wang2003,Noack2004,Mathew2007a,Finn2011}.

Quantification of mixing typically focuses on characterizing the amount of stirring by chaotic advection~\cite{Aref1984}, which in turn relates to the role that stretch-and-fold and stretch-and-stack mechanisms play in mixing of material~\cite{Childress1995,Wiggins2004,Sturman2006,Ottino1989,Villermaux2019}. The strength of these mechanisms is computed by estimating local rates of exponential stretching of the material~\cite{Thiffeault2001,Thiffeault2004} over a timescale associated with folding or stacking. Alternatively, chaotic advection can be quantified by studying norms of scalar fields advected in the flow~\cite{Mathew2005,Thiffeault2008a,Thiffeault2012}. A review of associated topics with additional references can be found in~\cite{Aref2017Rev.Mod.Phys.}.

The search for structures that enhance or prevent mixing typically focuses on features of the flow that remain coherent over relevant timescales and organize transport by, for example, attracting, repelling, or containing the material. The most well-known of such structures are the Lagrangian Coherent Structures (LCS)~\cite{Haller2015}, which are distinguished low-dimensional barriers to transport, and Almost-Coherent Sets~\cite{Froyland2009}, which are regions in which the transport is contained. In all cases, the true challenge remains to define the objects of interest so that they encompass all relevant physical phenomena, and to propose and implement an algorithm that identifies them, especially in transient or aperiodic flows \cite{Peacock2013, Samelson2013, Haller2015, Allshouse2015b, Hadjighasem2017,Balasuriya2018}.  The detection of coherent structures offers especially-significant applications to geophysical flows, where these techniques have been used to understand climate change and plan responses to ecological catastrophes~\cite{Allshouse2017EnvironFluidMech, Beron-Vera2015, Mezic2010, Tallapragada2011Chaos,BozorgMagham2013}.  The most commonly used tools to detect coherent structures are based on material deformation~\cite{Haller2011, Budisic2016c,Haller2015, Ma2015} and on probabilistic~\cite{Froyland2009,Froyland2012} properties of the flow. A comparison of the wide range of approaches for detecting coherent structures can be found in \cite{Hadjighasem2017}.

Input data in the form of continuous velocity field stands in contrast to observations in oceans and atmosphere that are recorded by deploying sparse sensors that measure physical properties (temperature, salinity, etc.) of the fluid flow as they are advected~\cite{Roemmich2009,Nodet2006,Coelho2015}. The relative sparsity of sensors prevents a reliable estimation of a continuous velocity field; to analyze such sparse datasets, a number of methods have been developed that require only a finite set of discretized trajectories~\cite{Allshouse2012,Froyland2015a,Hadjighasem2016Phys.Rev.E,Schlueter2017,Rypina2017,Williams2015a}.

Sparse-data methods include the \emph{braid dynamics} methods employed in this paper, which require only a set of discrete trajectories instead of the full velocity field. Following \citet{Boyland2000}, the study of the topological properties of fluid stirring has developed into an active research area.  The topological approach is particularly well-suited to the study of mixing by rods, vortices, or otherwise distinguished Lagrangian trajectories~\cite{Thiffeault2005,Thiffeault2006,Gouillart2006,Finn2007,Francois2015, Candelaresi2017Chaos,Taylor2016,Grover2012,Roberts2019}. In particular \citet{Thiffeault2005,Thiffeault2010} and \citet{Allshouse2012, Budisic2015} used braid theory to characterize mixing and coherent structures from planar flows solely from particle trajectories, forming the basis for the approaches used in this paper.

In planar flows, the rate of exponential growth of material interfaces corresponds to the topological entropy of the flow~\cite{Boyland1994,Boyland2000,Thiffeault2010,Thiffeault2005, Newhouse1993, Newhouse1988, Yomdin1987}, which quantifies the complexity of trajectories evolving in a dynamical system. Techniques for estimating  topological entropy (without estimating growth rate first) typically use precise
velocity fields to compute unstable periodic orbits \cite{Auerbach1987Phys.Rev.Lett.a,Davidchack2001PhysicsLettersA,Davidchack2000,Davidchack1999}, or intricate partitions of the flow~\cite{Froyland2001,Bollt2001PhysicaD:NonlinearPhenomena}.  \citet{Budisic2015} developed a method for calculating the Finite-Time Braiding Exponent (FTBE), which is an approximation for the topological entropy that is more applicable to finite-time, sparse datasets.  The FTBE provides a robust measure of mixing that approaches the topological entropy as the number of trajectories is increased. More recently, \citet{Roberts2019} developed a braid-free approach that also estimates topological entropy based on the relationship with growth of material lines, without detailed knowledge of the velocity field.

Since the methods of \citet{Budisic2015} and \citet{Allshouse2012} were developed and applied to model flows, the principal aim of this paper is to evaluate how well suited braid theory is for the characterization of experimental fluid flows. The flow studied here is the rotor-oscillator Stokes flow, a canonical example of a flow field that possesses both chaotic and coherent regions, described by \citet{Hackborn1997} and \citet{Weldon2008}.  In contrast to flows such as the double-gyre, Bickley jet, or the Duffing oscillator, which have been more commonly used to analyze material transport, the rotor-oscillator flow has been analyzed both as an analytical model~\cite{Hackborn1997} and as an experimental flow~\cite{Weldon2008}.  The braid-based analysis will be twofold: first, it will provide an identification of regions of the flow surrounded by \emph{minimally} growing material lines (typically, coherent material sets); second, an estimation of \emph{maximum} growth rate of material interfaces (typically found in the chaotic region). Both calculations will be applied to the analytical model and to the experimental data, in an effort to expose and address challenges that come with analyzing non-idealized flows.

The paper is organized as follows. Section~\ref{sec:roflow} presents the model and experimental flows used in this study. Section~\ref{sec:braids} gives a short summary of the braid representation of the flow kinematics. Next, the paper demonstrates how two particular analysis methods can be applied to model and experimental flows: Section~\ref{sec:results-lcs} explains the detection of coherent structures, comparing the analysis of numerically-advected and experimental particle trajectories, while Section~\ref{sec:results-mixing} explains estimation of material growth, comparing the analysis of model and experimental velocity fields.  Section~\ref{sec:discussion} summarizes what has been learned by employing braid dynamics to study model and experimental rotor-oscillator flows.

\section{The Rotor-Oscillator Flow}\label{sec:roflow}

The rotor-oscillator flow is a planar, non-autonomous, incompressible flow where the motion of the fluid is induced by a rotor that simultaneously rotates about its axis and oscillates along a channel.  Experimentally, particle motion is confined to a two-dimensional fluid layer in a rectangular tank, with advection induced by a fast-spinning cylinder that also oscillates in the longitudinal direction. This section summarizes the analytic model and the experimental realization; further details are found in~\cite{Hackborn1990,Hackborn1997,Weldon2008}.

\subsection{Analytic model}\label{sec:roflow-num}

\citet{Hackborn1997} studied the \emph{rotor-oscillator} flow, an asymmetric flow between parallel plates driven by the rotation of a cylinder (\emph{rotor}) and the longitudinal oscillation of one of the walls of the flow tank (\emph{oscillator}). In a suitable parameter regime, the flow exhibits a main vortex around the rotor, with two secondary recirculating vortices on each side of the rotor. Each of those vortices creates its own, progressively-weaker, array of vortices, which will not play a role in this work.

\begin{figure}
	\begin{centering}
		\includegraphics[width=0.45\textwidth]{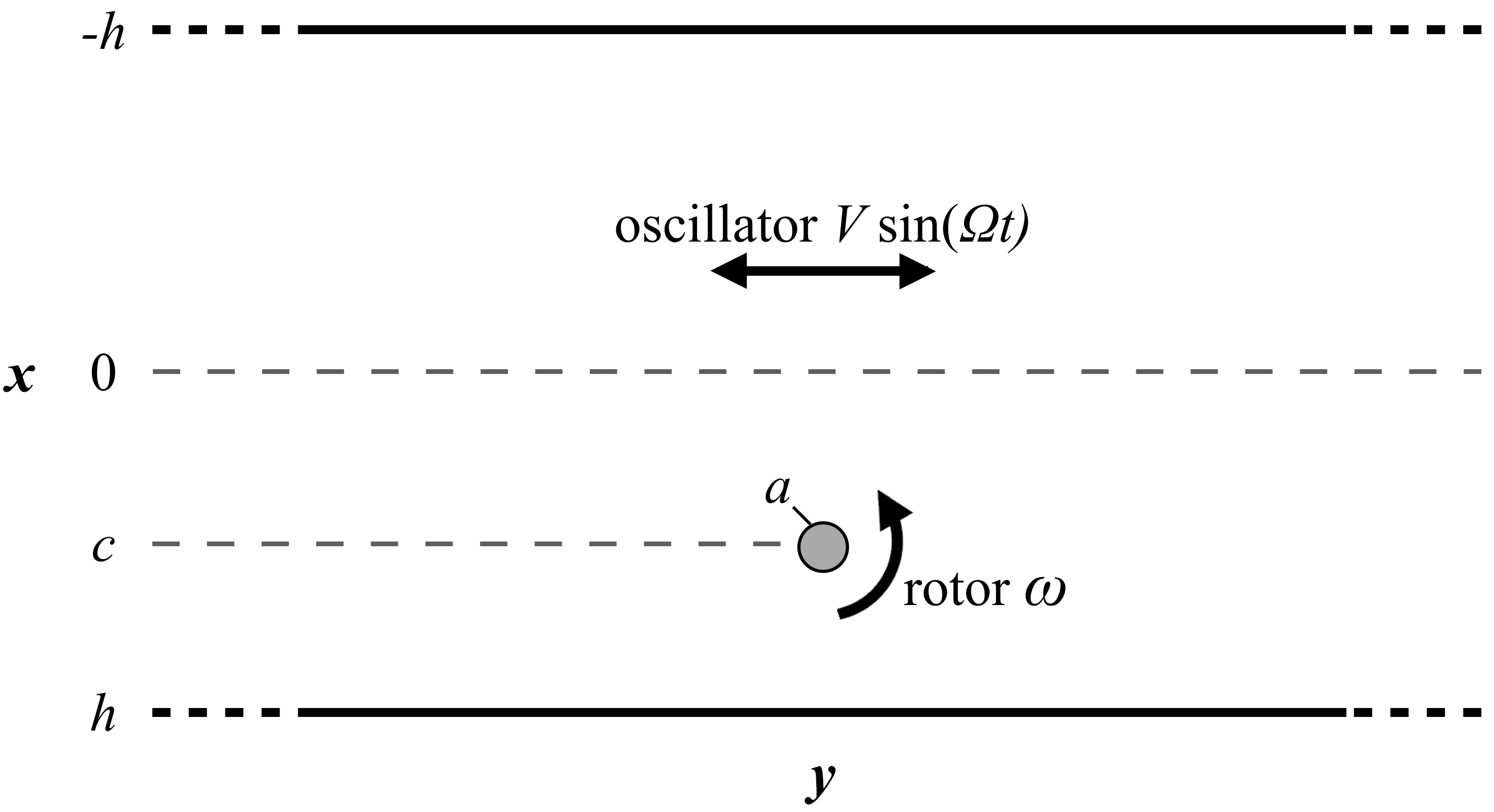}
		\caption{Geometry for the rotor-oscillator flow. The walls are at \(x = \pm h\), with the rotor oscillating in~\(y\) at constant~\(x=c\).}
		\label{fig:diagram}
	\end{centering}
\end{figure}

We take coordinates~$(x,y) \in [-h,h] \times \mathbb{R}$, with the walls of the channel at \(x=\pm h\). The cylindrical rotor starts at \(y=0\) at \(t=0\) and moves along the line \(x=c\). The rotor has radius \(a\) and angular velocity \(\omega\). A diagram of the geometry is shown in \autoref{fig:diagram}.  (The peculiar interchange of $x$ and $y$ axes is for consistency with earlier work \cite{Hackborn1997}.)
% fig:apparatus

Without longitudinal (\(y\)) oscillation, the rotating cylinder induces, in the Stokes limit, a steady velocity field. In non-dimensional coordinates, the span-wise width is \(x \in [-1,1]\) with the position of the rotor \(c = 0.54\) set to match~\cite{Hackborn1997}.  We absorb the rotation frequency \(\omega\) into the non-dimensional time \(t\), so the resulting stream function \(\psi(x,y)\)~\cite{Hackborn1990,Hackborn1997} is
\begin{equation}
  \psi(x,y) = \underbrace{\tfrac{1}{2}\log[f(x,y)]}_{\text{point vortex}} + \underbrace{\int _0^{\infty} g(x,k)\cos(ky)dk }_{\text{boundary effects}} \label{eq:steady-stream-function}
\end{equation}
where
\begin{equation*}
  \begin{aligned}
    f(x,y)
  &= \frac{\cosh(\pi y/2) - \cos(\pi(c-x)/2)}{\cosh(\pi y/2) + \cos(\pi(c+x)/2) }\,;\\
	g(x,k) & = \frac{2[\tanh k\cosh kx-x\sinh kx]\cosh kc}{\sinh 2k+2k} \\
		 & + \frac{2[\coth k\sinh kx-x\cosh kx]\sinh kc}{\sinh 2k-2k}.
\end{aligned}
\end{equation*}

Following~\cite{Hackborn1997}, we assume the rotor to be a point vortex with a vanishing diameter (\(a\rightarrow0\)) while keeping \(a^2\omega\) constant. The time non-dimensionalization is kept as follows:
\begin{equation}
	\frac{ta^2\omega}{h^2}\rightarrow t.%,\quad
	%\frac{\psi}{a^2\omega}\rightarrow \psi,\quad
	%\frac{\text{length}}{h}\rightarrow\text{length}
\end{equation}

The longitudinal oscillation of frequency \(\Omega\) is modeled by the time-periodic translation of the steady stream function \(\psi(x,y,t)\):
\begin{equation}
  \Psi(x,y,t) = \psi(x,y - \epsilon \sin \lambda t),
\end{equation}
with non-dimensional parameters
\begin{equation}
	\epsilon = \nofrac{Vh}{2a^2\omega}, \quad
	\lambda = \nofrac{h^2\Omega}{a^2\omega}, \label{eq:parameters}
\end{equation}
that combine the channel width \(h\), magnitude of oscillation \(V\) and angular frequencies of rotation \(\omega\) and of oscillation \(\Omega\).  The use of a translated steady solution is valid as long as~\(\Omega\) is not so large as to invalidate the Stokes approximation.

\begin{figure}
  % \graphicspath{ {./images/} }
  \begin{tikzpicture}
    \begin{axis}[y dir=reverse,
      axis on top,
      % xticklabel pos=top,
      xtick={-3,-2,...,3},
      xmin=-3,xmax=3,
      ymin=-1,ymax=1,
      axis equal image,
      grid=both,
      grid style={green!75!black,dashed},
      width=\linewidth,
      ylabel style={rotate=-90},
      xlabel={\(y\) },
      ylabel={\(x\) }
      ]
      \addplot[plot graphics/node/.append style={yscale=-1,anchor=north west}] graphics [xmin=-3,xmax=3,ymin=-1,ymax=1] {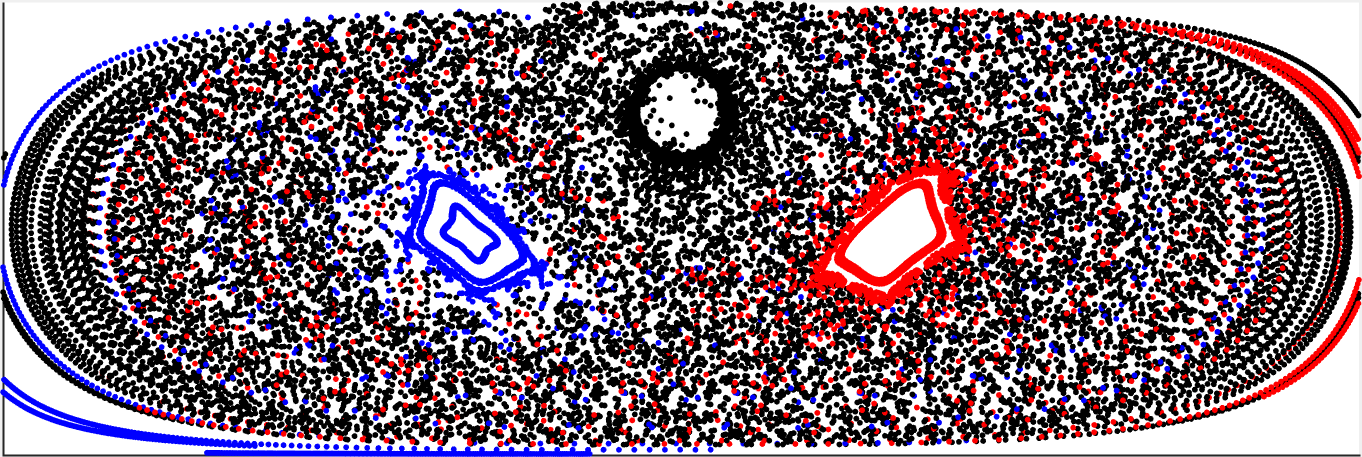};
      \draw[dashed,line width=1.5pt,color=green] (-1.5,-0.5) rectangle (1.5,0.5);
    \end{axis}
  \end{tikzpicture}

  \begin{tikzpicture}
    \begin{axis}[axis on top, y dir=reverse,
      xmin=-1.5,xmax=1.5,
      ymin=-0.5,ymax=0.5,
      ytick={-0.5,-0.25,...,0.5},
      axis equal image,
      grid=both,
      grid style={green!75!black,dashed},
      width=\linewidth,
      ylabel style={rotate=-90},
      xlabel={\(y\) },
      ylabel={\(x\) }
      ]
      \addplot[plot graphics/node/.append style={yscale=-1,anchor=north west}] graphics [xmin=-1.5,xmax=1.5,ymin=-0.5,ymax=0.5] {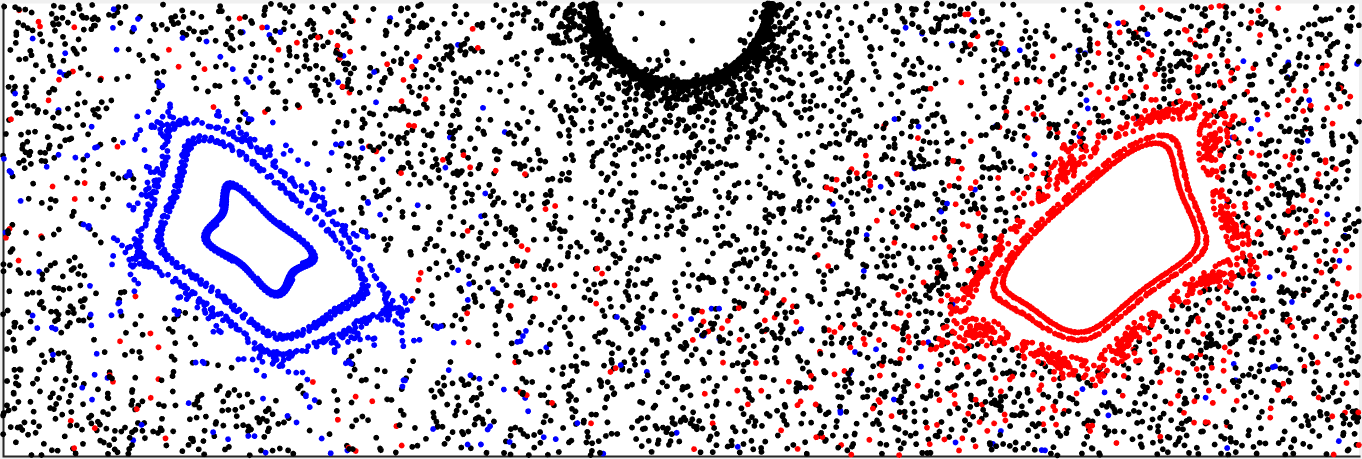};
    \end{axis}
  \end{tikzpicture}

	\caption{Poincar\'e map for the model rotor-oscillator flow \cite{Hackborn1997,Weldon2008} with  $c = 0.54$, $\epsilon = 0.125$, $\lambda = 2\pi/5$ simulated for 300 forcing cycles. The blue and red colors depend on the average value of the \(y\) coordinate along simulated trajectories. Bottom panel shows a detailed view of the dashed frame. The Poincar\'{e} map is sampled at the rotor's oscillation period \(\tau\).}
	\label{fig:poincare-map}
\end{figure}

Passive particles are advected via
\begin{equation}
  \begin{aligned}
  \dot x(t) &= -\partial_{y}\psi(x, y - \epsilon\sin\lambda t), \\
  \dot y(t) &= \phantom{-}\partial_{x}\psi(x,y - \epsilon\sin\lambda t).
\end{aligned}\label{eq:psi2}
\end{equation}
Fixing \(c = 0.54\), \citet{Hackborn1997} varied \(\epsilon\) and \(\lambda\) to identify regimes with chaotic dynamics and islands of coherence using Poincar\'e maps. \autoref{fig:poincare-map} shows a Poincar\'e map generated with $\epsilon = 0.125$ and $\lambda = {2\pi}/{5}$.

Timescales that govern the evolution of a single Lagrangian trajectory vary widely depending on its location. Trajectories that encounter the rotor will circle around it rapidly; trajectories that remain on the outside edge of both the rotor and recirculating vortices evolve on timescales separated by two or more orders of magnitude compared to the rotor rotation.

\subsection{Experimental flow} \label{sec:roflow-exp}

\begin{figure}
	\begin{centering}
	\subfigure[\ ]{ \centering
		%\graphicspath{ {./images/} }
			\includegraphics[width=0.475\textwidth]{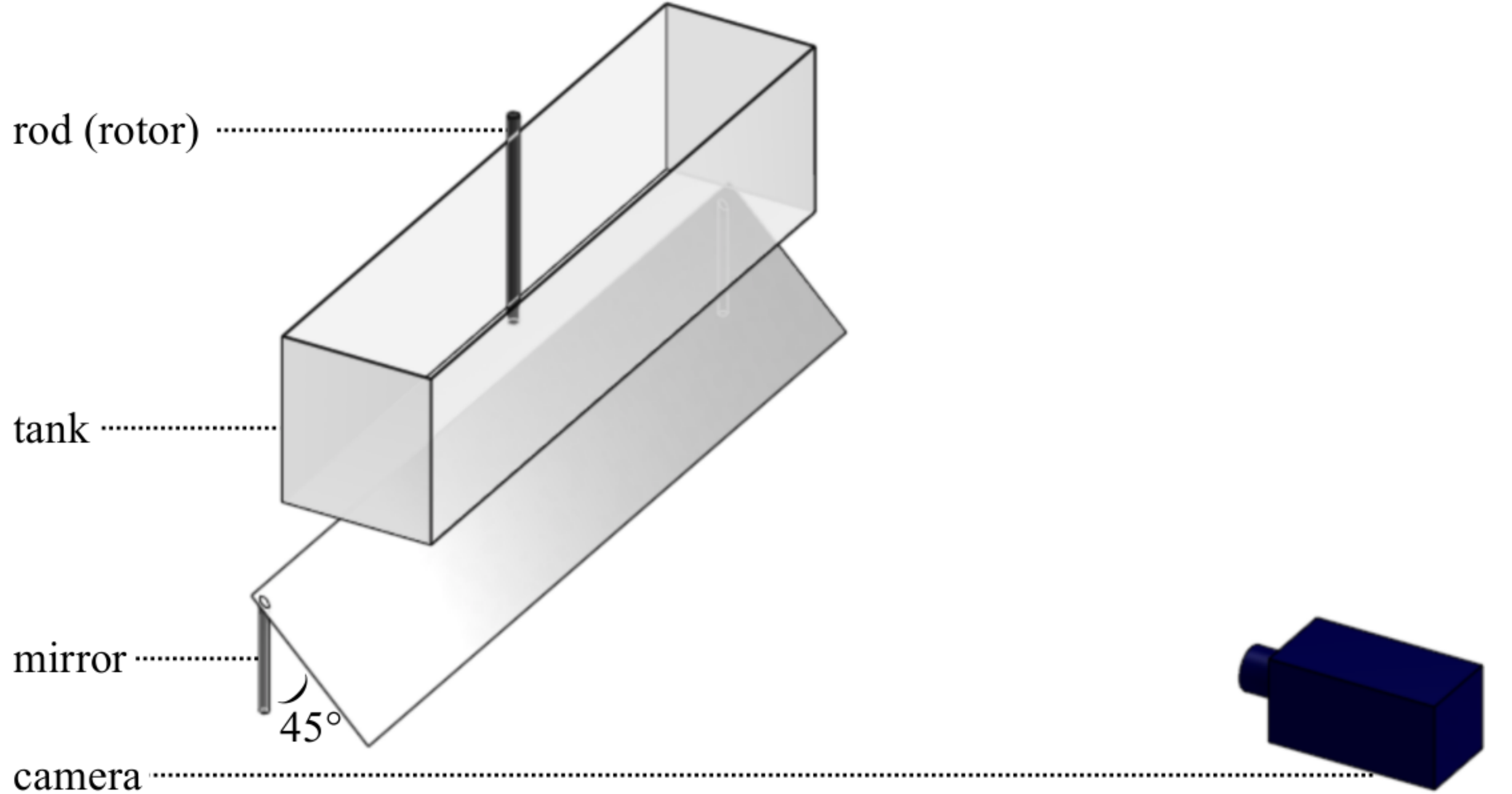}}
	\subfigure[\ ]{ \centering
			\includegraphics[width=0.475\textwidth]{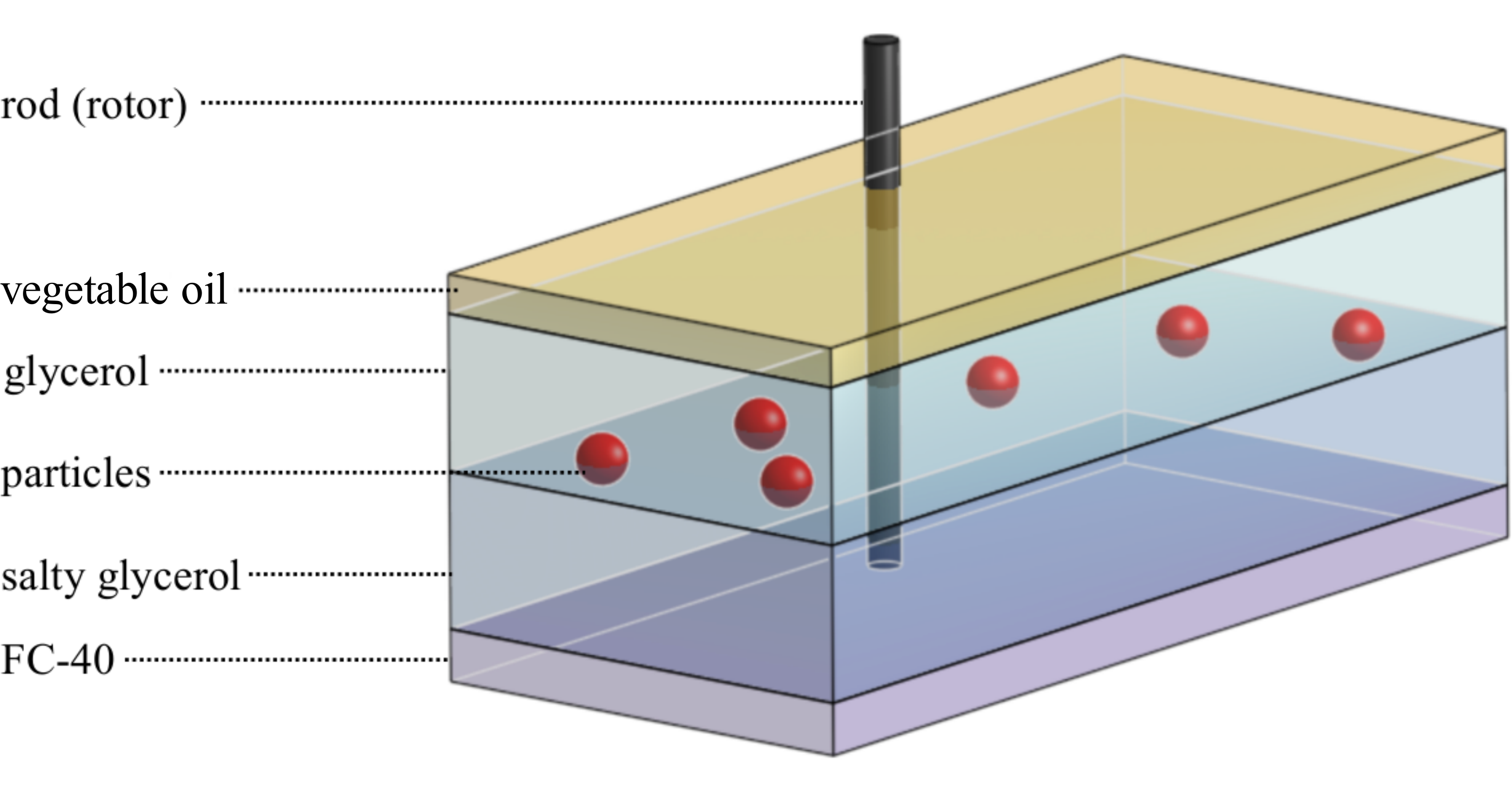}}
		\caption{Schematics of the experimental set-up for the Hackborn--Weldon rotor-oscillator flow \cite{Hackborn1997,Weldon2008}, with still walls and an oscillating rotor. (a) Experimental apparatus set up. (b) Fluid system to set up a two-dimensional flow through density layers for the PT experiment. In the PIV experiment, the tracer particles were mixed to a single layer of glycerol and a laser sheet was projected horizontally.
		 }
		\label{fig:apparatus}
	\end{centering}
\end{figure}

The experimental apparatus was inspired by \citet{Weldon2008}, and is depicted in \autoref{fig:apparatus}. The main difference compared to the analytical model  of \citet{Hackborn1997} is in apparent change of the reference frame: they fix the rotor and moved the walls, while \citet{Weldon2008} fix the walls and move the rotor.

The rotor-oscillator system was recreated in a $\SI{90 x 402}{\milli\meter}$ acrylic flow tank. For the rotor, an aluminum rod of diameter $a = \SI{3.165}{\milli\meter}$ was attached to a stepper motor through a plastic sleeve for thermal insulation.  To longitudinally oscillate the rotor, the stepper motor was mounted to a longitudinal traverse controlled by a second stepper motor.

The tank and the traverse were mounted onto aluminum frames and aligned horizontally. To visualize a horizontal cross-section of the system, a front-faced mirror was placed underneath the tank at a \ang{45} angle. A camera was mounted facing the mirror with the rod at the center of the field of view.  The experimental images were acquired through the mirror reflection at \ang{45} and recorded with a LaVision Imager Pro X 4M CCD camera of resolution \num{2042 x 2042} pixels with a $\SI{28}{\milli\meter}$ lens. Image acquisition, calibration and cross-correlation was performed by LaVision's DaVis imaging software.  To mitigate three-dimensional effects due to evaporation or interaction with the bottom wall, the fluid system was trapped between a top layer of vegetable oil and a bottom layer of FC-40 coolant.

Two separate experiments were used to record properties of the flow: a particle tracking experiment (PT, Section~\ref{sec:roflow-exp-PT}), to record Lagrangian trajectories, and a particle image velocimetry experiment (PIV, Section~\ref{sec:roflow-exp-piv}), to record instantaneous velocity fields.  Table \ref{tab:exp-parameters} summarizes the experimental parameters for the two experiments. For all experiments, the Reynolds number was estimated to be $\mathcal{O}(1)$.
% fig:apparatus

\begin{table*}[htb]
\caption{Experimental parameters.} % put the caption before the table
  \begin{threeparttable}
  \centering
  %alternate row colors before
  % center the headers
  \begin{tabular}[c]{c l l l l}%{| m{23pt} | m{35pt} | m{35pt}|}
  \hline
  \multicolumn{1}{c}{symbol} &
  \multicolumn{1}{c}{description} &
  \multicolumn{1}{c}{value (PIV)} &
  \multicolumn{1}{c}{value (PT)}\\\hline
    \(h\) & half-width of the channel& \multicolumn{2}{c}{$\SI{0.045}{\meter}$ }\\
    \(a\) & rotor radius& \multicolumn{2}{c}{$\SI{0.003165}{\meter}$ }\\
    \(c\) & distance of rotor from the wall& \multicolumn{2}{c}{$\SI{0.02376}{\meter}$ }\\
    \(\taudim\) & translational period of oscillation & $\SI{96.5201}{\second}$ & $\SI{94.7163}{\second}$  \\
    \(V\) & translational velocity magnitude & $\SI{5.8275e-4}{\meter}$ & $\SI{5.9385e-4}{\meter}$ \\
    \(\Omega\) & translational angular velocity& $\SI{0.0651}{\radian\per\second}$ & $\SI{0.0663}{\radian\per\second}$ \\
    \(\omega\) & rotational angular velocity& \multicolumn{2}{c}{$\SI{10.4720}{\radian\per\second}$ } \\
    \(\Re\) & Reynolds number\footnotemark[1]& \multicolumn{2}{c}{$\SI{2.6640}{}$ }\\
    \(f\) & sampling frequency & $\SI{1}{Hz}$\footnotemark[2] & $\SI{10}{Hz}$\\
    \(\epsilon\) & non-dim.\ oscillation magnitude $(Vh/2a^2\omega)$  & $0.1250$ & $0.1274$ \\
    \(\lambda\) & non-dim.\ oscillation ang.\ frequency $(h^2\Omega/\omega a^2)$ & $2\pi/5 \approx 1.2566$ & $1.2806$\\\hline
  \end{tabular}
      \begin{tablenotes}
      \small
            \item \footnotemark[1] In calculating the Reynolds number for the experiments, we use \(\omega a\) for the characteristic linear velocity and \(2h\) as the characteristic length, resulting in \(\Re=2\omega ah/\nu\). \citet{Hackborn1997} use
            \(a\) as the characteristic length, resulting in \(\Re=a^2\omega/\nu \approx \SI{9.3682e-2}{}\) here.
            \item \footnotemark[2] Effective frequency of sampling. See Section~\ref{sec:roflow-exp-piv} for details.
    \end{tablenotes}
    \end{threeparttable}
  \label{tab:exp-parameters}
\end{table*}

\subsubsection{Particle tracking (PT) experiment}
\label{sec:roflow-exp-PT}

Tracer particles were custom-made from cellulose acetate polymer spheres by Cospheric, with diameter $\SI{1.78}{\milli\meter}$ and density $\SI{1.285}{\gram\per\cubic\centi\meter}$. The particles were painted fluorescent to create a high light contrast and thereby enhance image acquisition. The coated particles were then filtered by density to ensure they remained on a virtually two-dimensional plane, the surface between the lower layer of salty glycerol, of density $\SI{1.297}{\gram\per\cubic\centi\metre}$, and the upper layer of pure glycerol, of density $\SI{1.261}{\gram\per\cubic\centi\metre}$. The denser glycerol solution was mixed with salt and water to match the viscosity of pure glycerol, measured to be $\nu = \SI{1.1197 e-3}{\square\meter\per\second}$.  The fluorescent particles were illuminated with an ultraviolet light in a dark room. With the bead diameter $D = \SI{1.78 e-3}{\meter}$, the viscosity $\nu$, a rod of radius $a = \SI{3.165}{\milli\meter}$ and angular velocity of $\omega = \SI{10.4720}{\radian\per\second}$, the Reynold number of the beads at the rotor's boundary was \(\Re_{\text{bead}} = \frac{(\omega a)D}{\nu} \approx \SI{5.2689e-2}{}\). The velocity being maximized at the rotor's boundary, \(\Re_{\text{bead}}\) is an upper bound for the flow field, so the neutrally buoyant beads are assumed to move as passive tracers.

The images were acquired at a frequency $f = \SI{10}{\hertz}$. There were $\num{77918}$ frames (time steps) recorded, for a total run duration of $\SI{7791.8}{\second}$, corresponding to about $82.3$ periods. The particle trajectories were obtained through a MATLAB package, the Tracking Code Repository~\cite{track}. For the particle tracking experiment, the non-dimensional parameters were $\epsilon=0.1274$ and $\lambda=1.2806$.  The physical period of oscillation was $\taudim = \SI{94.7163}{\second}$.

\subsubsection{Particle image velocimetry (PIV) experiment}
\label{sec:roflow-exp-piv}

Tracers for the PIV experiment were hollow glass spheres of mean diameter $\SI{8}-\SI{12}{\micro\metre}$ mixed with glycerol. The Reynolds number of the glass spheres at the rotor's boundary was \(\Re_{\text{sphere}} \approx \SI{1.9977e-4}{}\) and are assumed to move as passive tracers. A pulsed Nd:YAG laser was powered to illuminate the glycerol--glass spheres solution in a horizontal plane normal to the tank front wall. To reduce artifacts and noise in the velocity fields, the domain was partitioned into two for image processing in DaVis; this helped compensate for the wide range of velocities between the rod's rotation and the slower motion on the outer domain. First, the region around the rod was processed at $f = \SI{10}{\hertz}$ and bin-averaged in 10-image blocks; second, the outer domain was processed at $f = \SI{1}{\hertz}$ and matched to the bin centers of the first part. The resulting effective sampling rate was $f = \SI{1}{\hertz}$. There were $\num{10730}$ frames recorded, for a total run duration of $\SI{1073}{\second}$, corresponding to about $11.1$ periods. For the PIV experiment, the non-dimensional parameters were $\epsilon=0.1250$ and $\lambda=1.2566$.  The physical period of rotor oscillation was $\taudim = \SI{96.5201}{\second}$, which corresponds to~$5$ dimensionless time units.

\begin{figure}
  \subfigure[\ ]{ \centering
  \begin{tikzpicture}
    \begin{axis}[y dir=reverse,
      axis on top,
      % xticklabel pos=top,
      xtick={-2,-2,-1,...,2,2},
      xmin=-2,xmax=2,
      ymin=-1,ymax=1,
      axis equal image,
      width=\linewidth,
      ylabel style={rotate=-90},
      xlabel={\(y\) },
      ylabel={\(x\) }
      ]
      \addplot[plot graphics/node/.append style={yscale=-1,anchor=north west}] graphics [xmin=-2,xmax=2,ymin=-1,ymax=1] {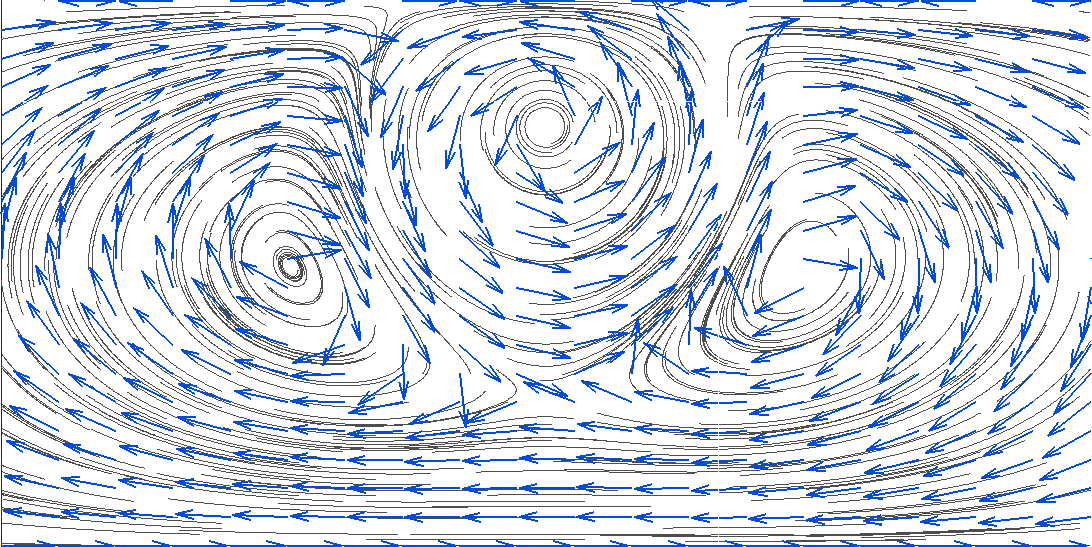};
    \end{axis}
  \end{tikzpicture}}
  \subfigure[\ ]{ \centering
  \begin{tikzpicture}
    \begin{axis}[y dir=reverse,
      axis on top,
      % xticklabel pos=top,
      xtick={-2,-2,-1,...,2,2},
      xmin=-2,xmax=2,
      ymin=-1,ymax=1,
      axis equal image,
      width=\linewidth,
      ylabel style={rotate=-90},
      xlabel={\(y\) },
      ylabel={\(x\) }
      ]
      \addplot[plot graphics/node/.append style={yscale=-1,anchor=north west}] graphics [xmin=-2,xmax=2,ymin=-1,ymax=1] {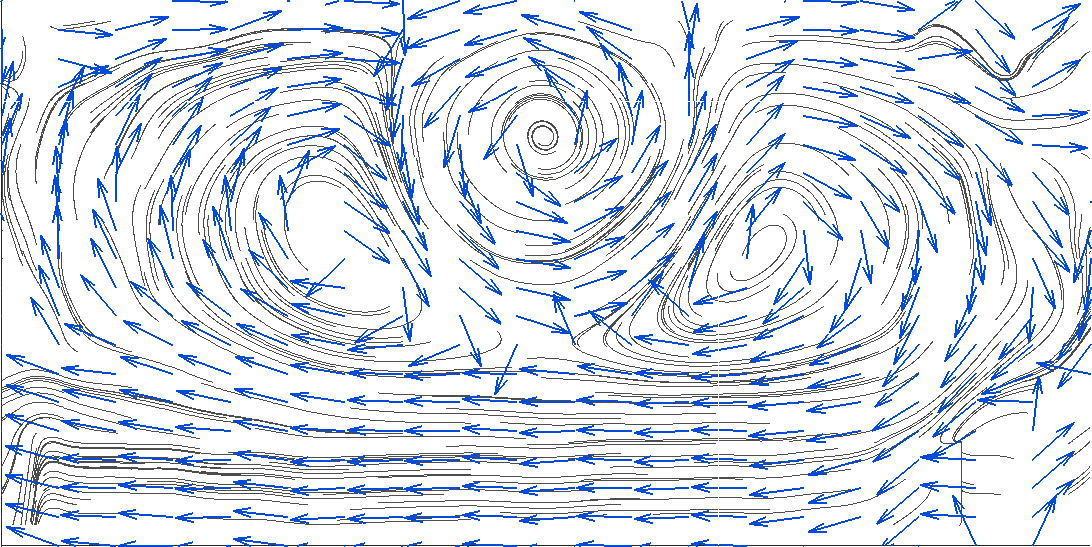};
    \end{axis}
  \end{tikzpicture}}
  \caption{Snapshot of instantaneous streamlines of (a) the model velocity field \cite{Hackborn1997,Weldon2008} with  $c = 0.54$, $\epsilon = 0.125$, $\lambda = 2\pi/5$ and (b) the PIV-recorded experimental velocity field at the same time instance.}
	\label{fig:streamlines}
\end{figure}

The velocity fields were post-processed using the approach of~\citet{Kelley2011PhysicsofFluids} to impose incompressibility for the PIV-obtained velocity field; this was necessary because of the multiple repeated periods over which the PIV data was used for our investigations. \autoref{fig:streamlines} demonstrates that instantaneous streamlines of the model and experimental velocity fields qualitatively match. Due to experimental constraints, however, the streamlines obtained from the PIV experiments deviate from the streamlines obtained from the model velocity fields. Some of these discrepancies come from the fundamental differences between the model and experimental flow. First, the model assumes the rotor to be a point vortex, whereas a physical rod had to be present to stir the fluid. In spite of the small rod diameter of $a = \SI{3.165}{\milli\meter}$, the flows deviate around the rotor. Second, the model assumes a Stokes flow with zero Reynolds number: with $a\neq 0$ as the characteristic length, the experimental Reynolds number is \(\Re \approx \SI{9.3682e-2}{}\), as described in Table \ref{tab:exp-parameters}. Additionally, the deviations are especially prominent near the lateral boundaries, due to the laser beam's reflection at the walls of the tank, which introduced artifacts in the velocity fields. The PIV record is not available for the full width of the channel, but rather in the approximate band \( -0.85 < x < 0.85\) in the nondimensional spanwise coordinate.
%in Figure 4. Is this because of the presence of side walls or is this the artifact of the PIV analysis?

To create synthetic trajectories in the experimental velocity field, the velocities were linearly interpolated between spatial nodes and time samples before passed on to the trajectory integrator. Trajectories were integrated using the variable-step, variable-order MATLAB \texttt{ode15s}~\cite{Shampine1999} algorithm to manage the stiffness of the differential equations, which is a consequence of the large difference in timescales of primary and secondary rotors. To simulate times longer than the recorded number of oscillation periods, the PIV velocity field was periodized in time.

\section{Braid dynamics}\label{sec:braids}

Braid theory is an algebraic way of characterizing and classifying continuous maps based on their topological properties. In our application, the continuous maps are the flow maps generated by the two-dimensional (unsteady) fluid velocity field over a particular time interval, as studied by \cite{Boyland2000,Boyland1994,Thiffeault2005,Thiffeault2006,Thiffeault2010}.  The ``input data'' for the braid theory characterization of fluid flows is a set of \(N\) continuous particle trajectories evolved concurrently over a finite time interval; in particular, the analysis does not require access to the velocity field or its gradients.

\begin{figure}[htb]
  \centering
  \subfigure[\ Physical braid]{ \centering
    \begin{tikzpicture}
      \begin{axis}[
        %view={0}{0}, %% x-projected
        smooth,
        width=.5\linewidth,
        height=3.25in,
        axis line style={draw=none},
        tick style={draw=none},
        xmin=-1.35,
        xticklabels={,,},
        yticklabels={,,},
        zticklabels={,,}]
        \addplot graphics [xmin=-1,xmax=1,ymin=-1,ymax=1] {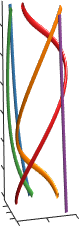};
        \node at (0.3,-1.1) {$x$};
        \node at (-1.2,-0.9) {$y$};
        \node at (-1.2,0) {$t$};
  \end{axis}
\end{tikzpicture}
    \label{fig:hackborn_braid_physical}
  }\hfill   \subfigure[\ Projection onto \(x\)-coordinate]{\centering
    \begin{tikzpicture}
      \begin{axis}[
        %view={0}{0}, %% x-projected
        smooth,
        width=.5\linewidth,
        height=3.25in,
        axis line style={draw=none},
        tick style={draw=none},
        xmin=-1.35,
        xticklabels={,,},
        yticklabels={,,},
        zticklabels={,,}]
        \addplot graphics [xmin=-1,xmax=1,ymin=-1,ymax=1] {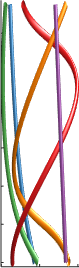};
        \node at (0,-1.15) {$x$};
        \node at (-1.2,0) {$t$};
  \end{axis}
\end{tikzpicture}
    }
  \hfill\subfigure[\ Topological braid]{
    \begin{tikzpicture}
  \braid[xscale=1,yscale=-1,%flip xy
  width=12pt, % spacing between strands
  style strands={1}{color=sRED, line width=2pt, line cap=butt},
  style strands={2}{color=sBLUE, line width=2pt, line cap=butt},
  style strands={3}{color=sGREEN, line width=2pt, line cap=butt},
  style strands={4}{color=sPURPLE, line width=2pt, line cap=butt},
  style strands={5}{color=sORANGE, line width=2pt, line cap=butt},
  height=15pt % size of element of the braid (crossing)
  ]
  % s_{4} s_{2}^{-1} s_{1} s_{3}^{-1} s_{2} s_{3} s_{4}^{-1} s_{2} s_{4} s_{3} s_{4}^{-1};
  s_{2}^{-1} s_{4} s_{1}
  s_{2} s_{3} s_{2}^{-1}
  s_{4}^{-1} s_{2} s_{4}
  s_{3} s_{4}^{-1};
\end{tikzpicture}
\label{fig:hackborn_braid_topological}
  }
  \caption[]{A physical braid and a corresponding topological braid generated from five trajectories.  % of the rotor-oscillator flow (see \S\ref{sec:roflow}).
In all diagrams the time flows from bottom to top. The sequence of generators (ordered from left to right in increasing time) is \( B =
  \sigma_{2}^{-1} \sigma_{4} \sigma_{1}
  \sigma_{2} \sigma_{3} \sigma_{2}^{-1}
  \sigma_{4}^{-1} \sigma_{2} \sigma_{4}
  \sigma_{3} \sigma_{4}^{-1}\).} \label{fig:sample-braid}
\end{figure}
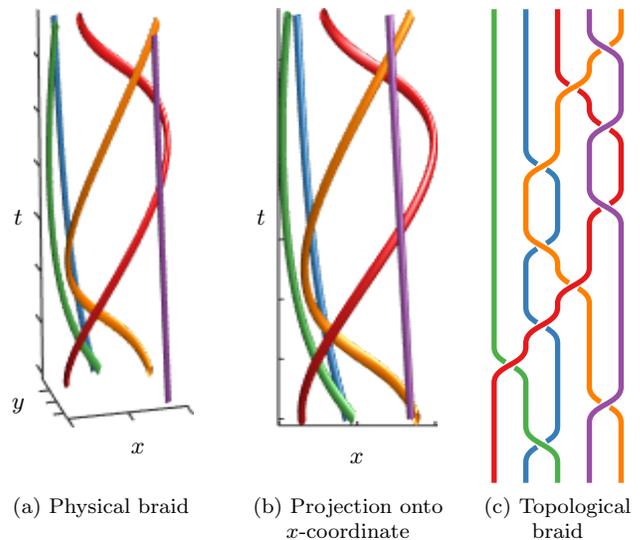

Braids are constructed from a set of \(N\) particle trajectories \(\mathbf{p}_{i}(t) \in \mathbb{R}^{2}\), \(i = 1,2,\dots,N\), with \(0 \le t \le T\).  A \emph{physical braid} is the embedding of trajectories in the three-dimensional space-time (Fig.~\ref{fig:hackborn_braid_physical}), where individual trajectories (\emph{strands}) weave around each other. A \emph{topological braid} (Fig.~\ref{fig:hackborn_braid_topological}) is a reduced representation of the physical braid that retains only the sequence of exchanges of the strand order with respect to a fixed space-time plane onto which the strands are projected. This plane can be chosen largely arbitrarily for the purposes of this article~\cite{Thiffeault2005,Thiffeault2010,Budisic2015,braidlab}.

Strand exchanges can be represented by a sequence of symbols \(\sigma_{i}\) called \emph{Artin braid generators}. The generator \(\sigma_{i}\) represents the crossing of the \(i\)th strand in front of
the \((i+1)\)th strand in Fig.~\ref{fig:hackborn_braid_topological}, where the index \(i\) indicates the order of strands from left to right just before the crossing occurs; the inverse generator \(\sigma_{i}^{-1}\) represents the crossing of \(i\)th strand behind the \((i+1)\)th strand.

\begin{figure}%[htb]
  \centering
  \subfigure[]{
    \includegraphics[width=0.42\linewidth]{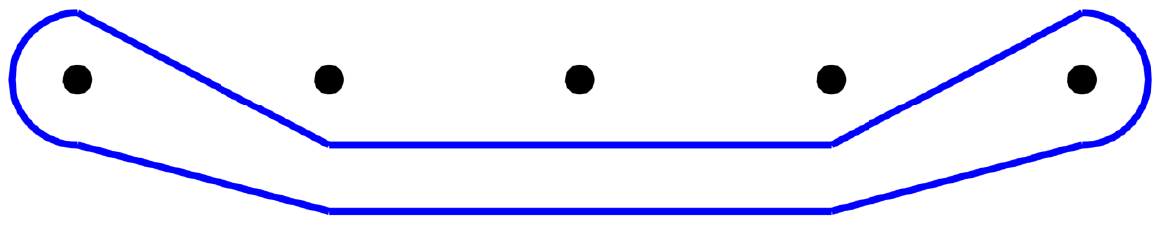}
\quad\vline\quad
    \includegraphics[width=0.42\linewidth]{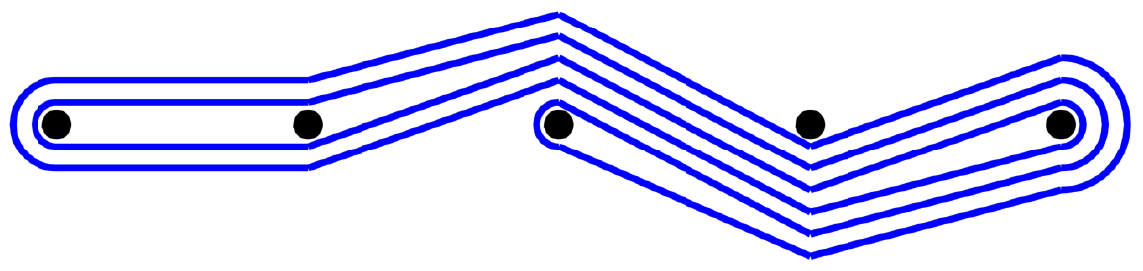}
    \label{fig:hackborn_loop1}
  }
  \subfigure[]{
    \includegraphics[width=0.42\linewidth]{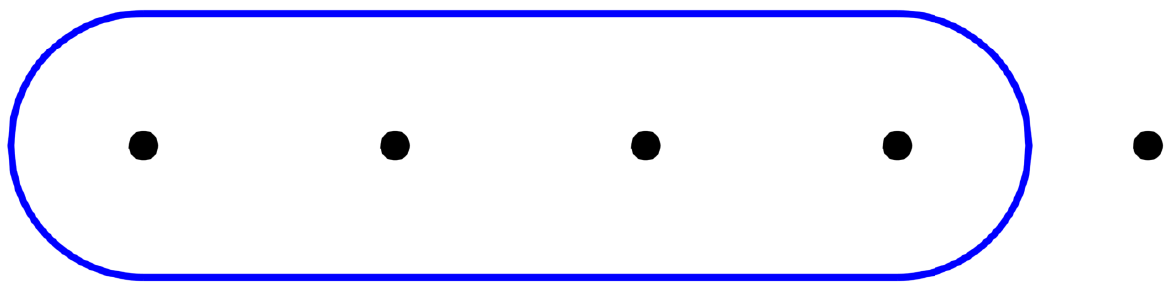}
\quad\vline\quad
    \includegraphics[width=0.42\linewidth]{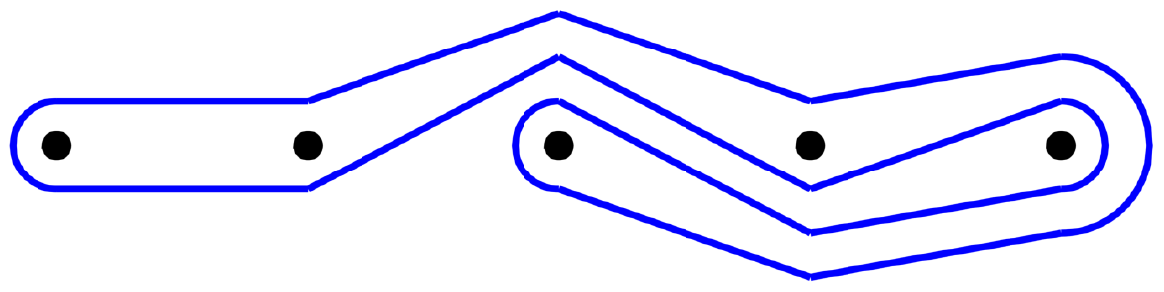}
    \label{fig:hackborn_loop2}
  }
  \caption[]{Two topological loops before (left column) and after (right column) action of the braid from \autoref{fig:sample-braid}. The strands of the braid are shown in black as cross-sections. The Dynnikov vectors before and after the braid action in \subref{fig:hackborn_loop1} are \(
  \begin{bsmallmatrix}
    1 & 1 & 1 & 0 & 0 & 0
  \end{bsmallmatrix}
  \xrightarrow{B}
  \begin{bsmallmatrix}
    0 & -2 & 3 & 0 & -1 & 0
  \end{bsmallmatrix}
  \) and in \subref{fig:hackborn_loop2}
\(
  \begin{bsmallmatrix}
    0 & 0 & 0 & 0 & 0 & 1
  \end{bsmallmatrix}
  \xrightarrow{B}
  \begin{bsmallmatrix}
    0 & -1 & 1 & 0 & -1 & 0
  \end{bsmallmatrix}
  \).}
\label{fig:sample-loops}
\end{figure}

A topological braid constructed from a given set of trajectories captures topological information about the flow. The topological analog of material advection in fluid flows is the action of a braid on \emph{topological loops}. A topological loop is a collection of closed, non-intersecting curves that enclose two or more (but not all) strands in the braid (\autoref{fig:sample-loops}).  The curves in a loop are ``pulled tight,'' i.e., they can be pictured as rubber bands tightly wrapped around strands of the physical braid. The action of a braid \(B\) on a loop \(\ell\), denoted by \(B\ell\), is visualized by sliding the rubber bands along the physical braid in the direction of time. As strands exchange places, the loop is forced to stretch since it cannot pass through the strands (\autoref{fig:sample-loops}).

\begin{figure}
  \centering
  \begin{tikzpicture}[scale=0.9, every node/.style={scale=0.9}]
    \node[inner sep=0pt] (loop) at (0,0)
    {\includegraphics[width=.5\textwidth]{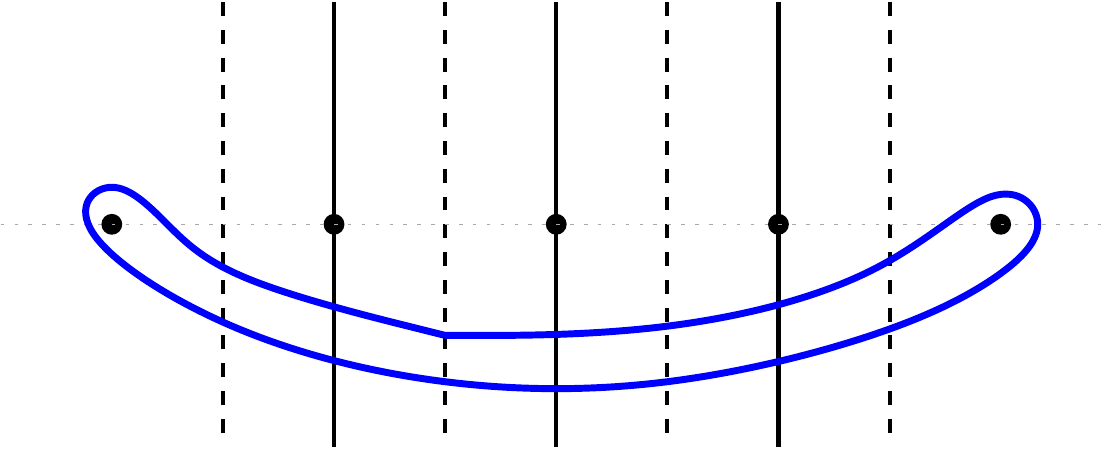}};
    \foreach \n in {-3,-1,1,3}
    { \pgfmathtruncatemacro{\label}{(\n+1)/2 + 2}
      \node at (\n,2) {\(\beta_{\label}\)};
    }
    \foreach \n in {-2,0,2}
    {
      \pgfmathtruncatemacro{\labeltop}{\n+3}
      \pgfmathtruncatemacro{\labelbottom}{\n+4}
      \node at (\n,2) {\(\alpha_{\labeltop}\)};
      \node at (\n,-2) {\(\alpha_{\labelbottom}\)};
    }

    \foreach \n in {-4,-2,0,2,4}
    {
      \pgfmathtruncatemacro{\label}{\n/2+3}
      \node at (\n,0.25) {\(\label\)};
    }

  \end{tikzpicture}
  \caption[]{Coordinate system based on intersection numbers \(\alpha_{i}, \beta_{i}\) used to calculate Dynnikov coordinates~\eqref{eq:dynnikov-vector}. In this example of \(N=5\) strands, \(\alpha_{1,3,5}  = 0\), \(\alpha_{2,4,6} = 2\), \(\beta_{1,2,3,4} = 2\), resulting in the Dynnikov vector  \( \begin{bsmallmatrix}
    1 & 1 & 1 & 0 & 0 & 0
  \end{bsmallmatrix} \). Precise location of strands in the \(x,y\) space is not important, only their order along the projection coordinate. }
  \label{fig:dynnikov-coordinates}
\end{figure}
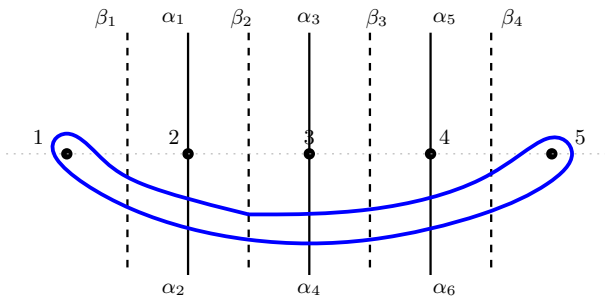

Computationally, any loop acted on by the \(N\)-strand braid can be represented by a vector of~$2N-4$ signed integers, and vice-versa, through the Dynnikov coordinatization of loops~\cite{Dynnikov2002,Hall2009,Thiffeault2010}. \autoref{fig:dynnikov-coordinates} demonstrates how to calculate the Dynnokov coordinates of a loop around \(N=5\) strands. The \emph{intersection numbers} \(\alpha_{1},\alpha_{3}, \cdots,\alpha_{2N-3}\) count the number of intersections between the tightened loop and axes above the punctures, while  \(\alpha_{2},\alpha_{4}, \cdots,\alpha_{2N-4}\) and \(\beta_{1},\cdots,\beta_{N-1}\) count the number of intersections with axes below the punctures and between the punctures, respectively. The Dynnikov coordinate vector is calculated from intersection numbers as
\begin{equation}
  \label{eq:dynnikov-vector}
  \begin{aligned}
  \begin{bmatrix}
    a_{1} & a_{2} & \cdots & a_{N-2} & b_{1} & b_{2} & \cdots & b_{N-2}
  \end{bmatrix}, \text{ where } \\
  a_{n} = (\alpha_{2n}-\alpha_{2n-1})/2, \quad
  b_{n} = (\beta_{n+1}-\beta_{n})/2.
\end{aligned}
\end{equation}

The action of each braid generator \(\sigma_{i}\) is then represented by a piecewise-linear map \(\sigma_{i}:\mathbb{Z}^{2N-4} \to \mathbb{Z}^{2N-4}\);  explicit expressions for \(\sigma_{i}\) can be found in~\cite{Hall2009,Thiffeault2010}. All braid computations and visualizations in this paper have been produced using the MATLAB toolbox \braidlab~\cite{braidlab}.

To measure the amount of stretching, we compute the ``length'' of a topological loop \(\abs{\ell}\) as the number of times it intersects the horizontal axis that would pass through all points in~\autoref{fig:sample-loops}. Although the quantity \(\abs{\ell}\) on its own does not correspond to spatial length of a material curve, it is useful for characterizing the relative loop growth factor
\begin{equation}
  \label{eq:loop-growth}
  \nofrac{\abs{B\ell}}{\abs{\ell}},
\end{equation}
measuring the ratio of loop length before and after application of the braid \(B\). The loop growth factor will play a prominent role in the following sections as a criterion for searching for maximally and minimally growing loops \(\ell\), given a specific braid \(B\).

The topological braid can be interpreted as a dramatically reduced representation of the full flow using only a finite number of trajectories, with the physical trajectories represented by topological operations, and material advection represented by stretching of ``rubber-band'' loops. Despite their simplicity, braid-based calculations provide bounds on the rate of material stretching in the full flow, and approximate boundaries of Lagrangian coherent structures, as explained in Sections~\ref{sec:results-lcs},~\ref{sec:results-mixing}.  Additionally, the reduction also yields significant speed up as compared to material line advection, which can rely on expensive interpolation methods~\cite{You1991}.

\section{Detection of coherent structures}\label{sec:results-lcs}

The braid theory approach to detect coherent structures offers a unique mechanism for partitioning different regions of the flow based on the entanglement of trajectories, as opposed to their relative positions~\cite{Froyland2015a} and material deformation~\cite{Haller2001a}.  A coherent structure identified using braids is defined as a topological loop that does not grow under the action of the braid.  The loop provides an approximate physical boundary that delineates an internal fluid subdomain that does not mix with exterior fluid.  We present an overview of the theory developed by \citet{Allshouse2012}, and show how the current work improves the algorithm to more efficiently detect coherent structures for larger trajectory sets.  The method is then applied to a synthetic set of trajectories based on the model velocity field presented in~\ref{sec:roflow-num} and on  experimentally measured trajectories, allowing for comparison between theory and experiments.

\subsection{Theory}
\label{sec:algorithm-coherent-structures}

There are several different definitions of coherent structures used for characterizing patches of passive tracer that do not disperse under Lagrangian transport~\cite{Haller2015,Froyland2010a,Ma2015}. In the context of braid theory, \citet{Allshouse2012} define coherent sets as a set of trajectories surrounded by an initially ``simple'' topological loop that does not grow, or grows sub-exponentially, over the duration of the braid.  Sub-exponential growth can occur as particles rotate as a patch, so we do not consider this type of growth as indicative of long term mixing.  We now summarize the relevant techniques from~\cite{Allshouse2012} with modifications to make the presentation more compact.

\begin{figure}
%\graphicspath{ {./images/} }
	\includegraphics[width=0.9\linewidth]{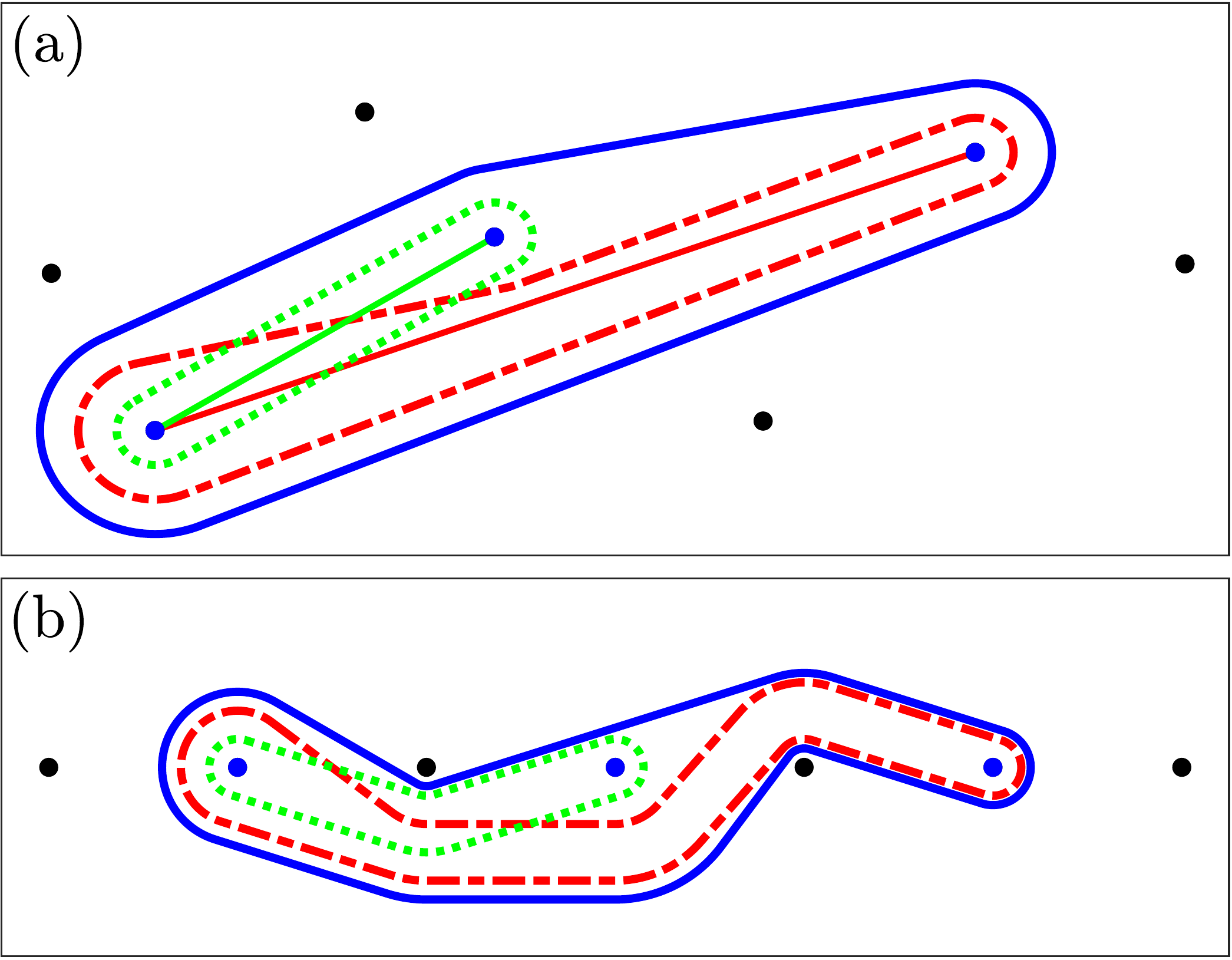}
	\caption[]{(a) Particles (dots) and loops in physical space.  Particles in the identified structure (blue) are connected by straight lines (solid) and surrounded by the corresponding straight pair-loop (dashed and dotted lines).  The coherent structure loop (blue line) links the two straight pair-loops.  (b) The particles and loops in standard form, with points aligned on the real axis.
	The Dynnikov vectors are
	\(
	\begin{bsmallmatrix}
	0 & 1 & 0 & -1 & 0 & -1 & 0 & 0 & 0 & 1
	\end{bsmallmatrix} % 	[0,1,0,-1,0,-1,0,0,0,1]
	\) for the blue curve,
	\(
	\begin{bsmallmatrix}
	0 & 1 & 1 & -1 & 0 & -1 & 0 & 0 & 0 & 1
	\end{bsmallmatrix} % 	[0,1,0,-1,0,-1,0,0,0,1]
	\) for the red curve and
	\(
	\begin{bsmallmatrix}
	0 & 1 & 0 & 0 & 0 & -1 & 0 & 1 & 0 & 0
	\end{bsmallmatrix} % 	[0,1,0,-1,0,-1,0,0,0,1]
	\) for the green curve.
	}
	\label{fig:straightLoops}
\end{figure}

Simple topological loops are those whose Dynnikov coordinates (see Section~\ref{sec:braids}) contain only \(\{\pm 1,0\}\), i.e., their coordinate vectors are \(\{\pm 1,0\}^{2N-4} \subset \mathbb{Z}^{2N-4}\).  As the magnitude of elements in the coordinate vectors relate to the number of folds of a loop around particles, these loops will have one or no folds as they surround particles (see figure~\autoref{fig:straightLoops} for examples). As mentioned in~\cite{Allshouse2012}, enumerating all simple loops and checking how fast they grow under the action of the braid is computationally intractable, even for modestly-size braids: \(\#( \{\pm 1,0\}^{2N-4} ) = 3^{2N - 4}\sim 9^{N}\), a number that is prohibitively large for even \(N \approx 20\) strands in a braid. Therefore, we look to reduce the number of simple loops considered, while still retaining a way to determine which advected particles are in coherent structures.

To reduce the number of loops considered, \citet{Allshouse2012} focuses on pair-loops that enclose only two out of the \(N\) particles. Although there are \(\binom{N}{2} = N(N-1)/2\) particle pairs to choose, there are still many simple loops that enclose any given pair. In contrast to  \cite{Allshouse2012} where loops were completely above or below all particles between the particle pair, here we consider a single pair-loop for each pair, namely the one that can be ``pinched'' to a straight line in the initial conditions plane; we term these \emph{straight pair-loops}.  This creates a set of $\binom{N}{2}$ loops to consider, which is half as many as in \cite{Allshouse2012}, while also yielding loops that have a simpler spatial shapes.  \autoref{fig:straightLoops}(a) contains the lines (red and green) connecting pairs of particles, and the corresponding straight pair-loops in standard form are presented in~\autoref{fig:straightLoops}(b).

The identification of coherent structures as loops that do not grow under the action of the braid is broken into two parts: identification of sets of particles enclosed by a coherent structure, and creation of a loop around those particles that does not grow.  The starting point for identifying sets of particles that are inside a coherent structure is the forming of straight pair-loops \(\ell_{ij}\) connecting particles \(\mathbf{p}_{i}, \mathbf{p}_{j} \in \mathbb{R}^{2}\) in the initial condition plane. We represent the collection of \(N\) trajectories by vertices in (a family of) undirected graphs \(\mathcal{G}\); the set of particles enclosed by a coherent structures will correspond to connected components in these graphs. The growth of the straight pair loops under the action of the braid will determine which edges \(e_{ij}\) are included or excluded in \(\mathcal{G}\).

Based on the loop growth rate \eqref{eq:loop-growth}, compute \(L_{ij}\) as the exponential growth rate of the straight pair-loop \(\ell_{ij}\) under the action of the braid
\begin{equation}
  \label{eq:pair-loop-growth}
  L_{ij} = \frac{1}{T} \log \frac{\abs{B\ell_{ij}}}{\abs{\ell_{ij}}}.
\end{equation}

The graph of vertices \(\mathcal{G}\) is fully-connected and undirected as there is a pair-loop between each pair of vertices, with edge weight \(L_{ij}\). Fix a threshold \(\Lambda \geq 0\) and construct a family of subgraphs \(\mathcal{G}_{\Lambda} \subset \mathcal{G}\), retaining only edges that satisfy \(L_{ij} \leq \Lambda\).  \(\mathcal{G}_{\Lambda}\) can be interpreted as a subgraph of \(\mathcal{G}\) where connections are made only by simple straight pair-loops whose exponential growth rate is less than \(\Lambda\).

For any \(\Lambda\) there may be vertices in \(\mathcal{G}_{\Lambda}\) that remain disconnected from any other vertex. In particular, vertices that are disconnected at \(\Lambda=0\) are deemed \emph{incoherent}, as they are not enclosed by any slow-growing pair-loops. On the other hand, \emph{coherent structures} will correspond to vertices that are mutually connected (connected components of the graph). Particles are grouped into a coherent structure by greedy agglomeration of the connected vertices.  We expect that, as \(\Lambda\) is increased from its lowest value, the number of coherent structures initially grows as more pair-loops become admissible. Eventually, additional pair-loops that are added will act as connections between pre-existing connected components, merging pairs of coherent structures. This will result in a decrease in the number of coherent structures, and growth in size of individual coherent structures, until there is only one large structure containing all particles within the system.

Once a set of particles has been identified as being contained by a coherent structure, the corresponding structure is the composite of the straight pair-loops that connect the set of particles.  In the example presented in \autoref{fig:straightLoops}, three particles (colored in blue) are connected by straight pair-loops, a green loop and a red loop.  The final step in the coherent structure identification is to merge the non-growing straight pair-loops into a single non-straight loop, which we consider to be the boundary of the coherent structure.  In \autoref{fig:straightLoops}, the coherent structure surrounds both straight pair-loops without enveloping any of the particles outside the coherent structure.

\subsection{Coherent structures in the model system}

While the braid method for detecting coherent structures was validated by~\citet{Allshouse2012}, the use of only straight pair-loops is a modification that requires similar verification.  To do so, we analyze a set of trajectories generated using the model velocity field presented in section~\ref{sec:roflow-num}.  Because only a single pair-loop per pair of trajectories is analyzed, it is possible to process a larger number of trajectories.  Increasing the number of trajectories while keeping the domain size constant has the potential to more accurately locate the boundary of the coherent set with the braid method.  As demonstrated by~\citet{Allshouse2012}, the braid method requires multiple particles to be located within a coherent structure in order for it to be identified; by increasing the number of trajectories analyzed, it is thus more likely that smaller structures will be identified. For the numerically calculated trajectories, 300 trajectories are analyzed and the resulting coherent structures can be compared with the Poincar\'e map in \autoref{fig:poincare-map}.

The numerically calculated trajectories are randomly distributed throughout the entire domain and are advected for 50 periods to match the PT experiment. A sampling rate of 100 measurements per period is sufficient to properly represent the braid. The 300-trajectory braid is represented by approximately 2.15 million generators. The application of the braid to all of the straight pair-loops is the bottleneck of the algorithm and takes three hours on a laptop. We use a growth rate threshold of $L_{ij} = 1/T$ to identify the straight pair-loops that do not grow rapidly over the time interval.  Using this growth rate, we identify loops that grow by a factor of $e^1$ over the time interval (see \eqref{eq:pair-loop-growth}), which is less than $1\%$ of all the pair-loops.

\begin{figure}
%\graphicspath{ {./images/} }
	%\includegraphics[width=0.9\linewidth]{CS_analyticLoops}
	\includegraphics[width=\linewidth]{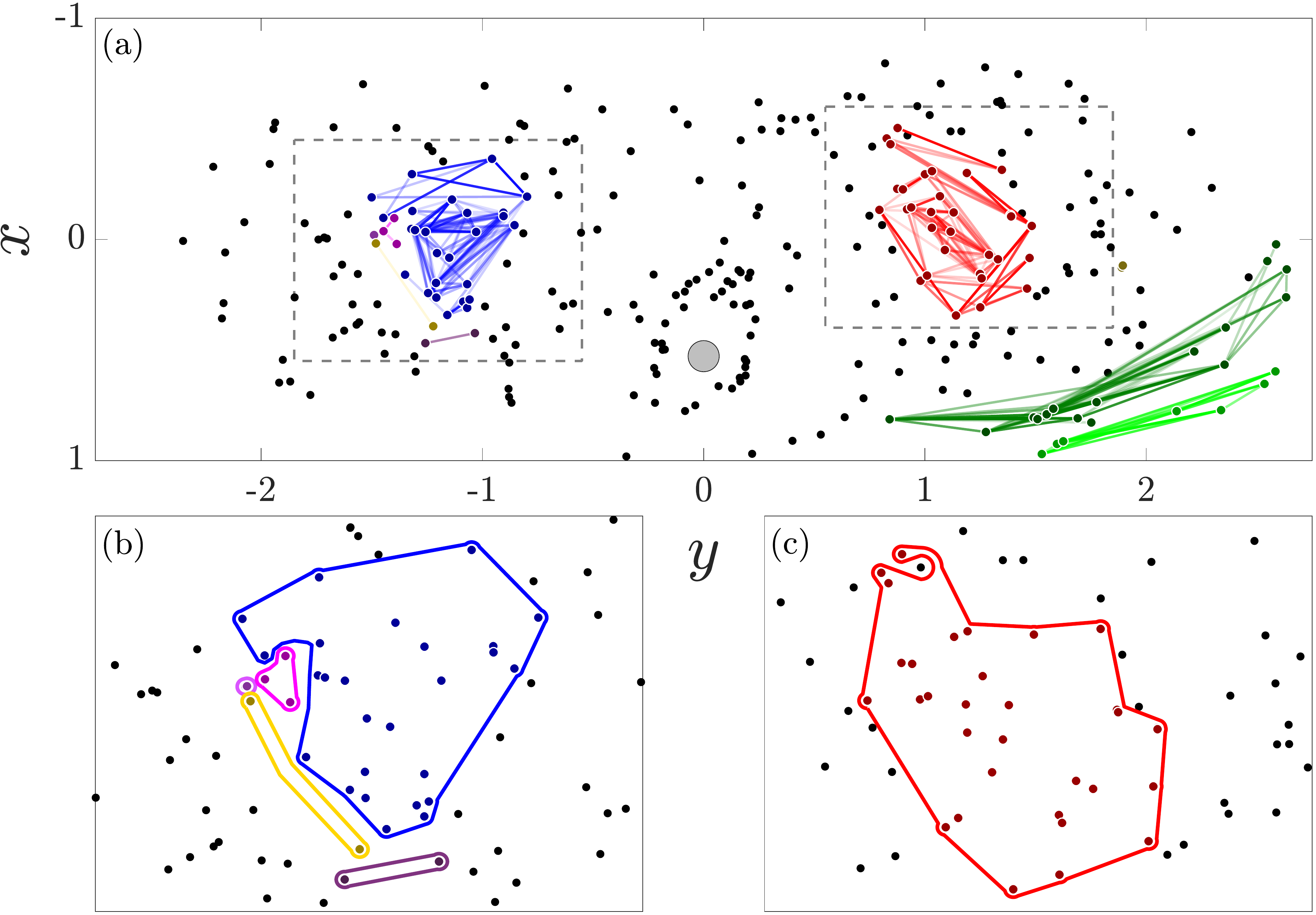}
	\caption{(a) The initial positions of \(N=300\) analytic model trajectories (dots), colored based on their coherent structure assignment with unassigned particles in black. The lines connecting particles indicate straight pair-loops that have a growth ratio less than the threshold, with less opaque lines representing loops that grow more. (b) and (c) are zoomed in regions corresponding to the gray dashed boxes in (a). The coherent structures corresponding to slowly growing loops for each set of points is drawn in the corresponding color.}
	\label{fig:analyticLoops}
\end{figure}

The network of particles connected by a slow-growing, straight pair-loop is presented as lines in~\autoref{fig:analyticLoops}(a). The particles that are connected by a network of these pair-loops form a coherent set, where each set has a different coloring of both the particles and the network connections. Finally, the coherent structure, which is represented by the slow-growing loop that surrounds all the particles of a particular coherent set, can be identified.  These resulting slow-growing loops are presented in~\autoref{fig:analyticLoops}(b) and (c).

Four different types of coherent structures were identified by the method.  The two structures in red and blue represent each of the main recirculating vortices.  Over the time interval, the topological length of the left and the right recirculating vortex loops grow by a factor of 1.24 and 1.48, respectively. Near the left vortex (blue), there are four additional structures identified at its perimeter: these structures are plotted in magenta, purple, yellow and dark purple. Each of these features is comprised of two or three particles and in all cases, the structures closely orbit the recirculating vortex without penetrating the boundary.  Another type of coherent structure identified is presented in light and dark green in~\autoref{fig:analyticLoops}(a).  These two structures correspond to particles that orbit around the chaotic sea and the recirculating vortices.  The dark green structure grows by a factor of 1.47 while the light green structure shrinks to 0.6 times its initial length due to the particles becoming closer in the topological projection.  Finally, the fourth type of structure is identified in gray in~\autoref{fig:analyticLoops}(a) near (1.89,0.13) cm.  This structure is made up of two particles that are a physical distance of 0.01 cm apart.  Because of their close proximity, these particles do not separate over the 50 periods of advection, despite being in the chaotic sea.

\subsection{Coherent structures in the experimental system}

Having identified the main coherent structures in the model system, we next apply the method to experimentally observed trajectories. The particle trajectories generated by the experimental system (see section~\ref{sec:roflow-exp-PT} and table \ref{tab:exp-parameters}) can be applied to the coherent structure method to identify the structures in the flow field.  The first 50 periods of advection are considered, throughout which 33 trajectories remained in the imaging domain for the duration of the time interval. These 33 trajectories produce a braid made up of 9,592 generators. Using the same length ratio threshold as the analytic model trajectory analysis, we identified three coherent sets of particles.  Each of the sets is surrounded by a non-growing loop. Applying the method to this dataset takes less than 10 seconds.

\begin{figure}
%\graphicspath{ {./images/} }
	%\includegraphics[width=0.9\linewidth]{CS_experimentalLoops}
	\includegraphics[width=.77\linewidth]{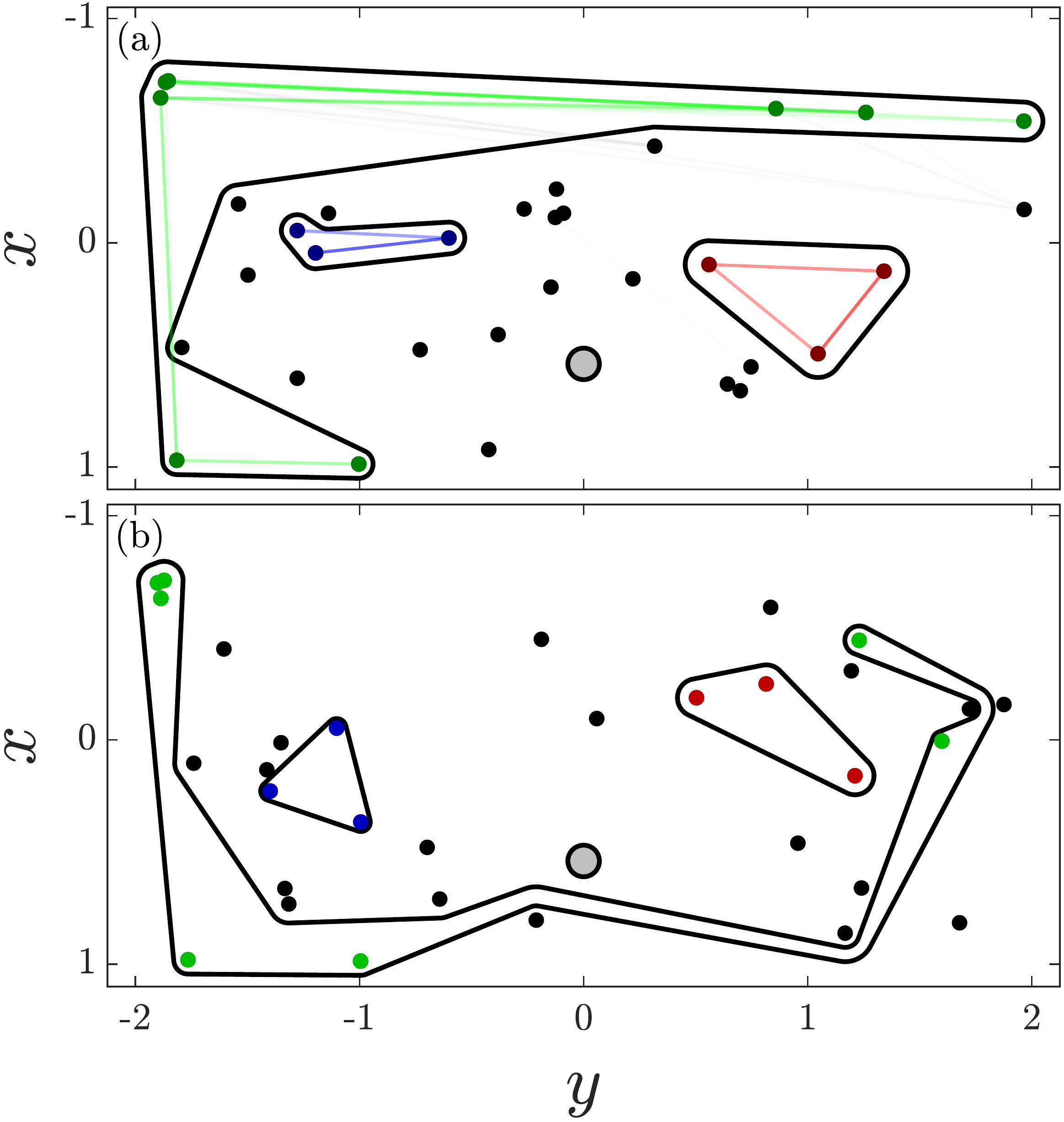}
	\caption{(a) The initial positions of \(N=33\) experimental trajectories (dots), and identified coherent loops (black lines).  Three coherent structures are identified (red, blue and green). The straight pair-loops that have a growth rate of less than the threshold are represented as straight lines connecting the particles. The less opaque the line, the greater the growth rate. (b) The final position of the trajectories and the deformed coherent loops.}
	\label{fig:expLoops}
\end{figure}

The initial and final position of the particles and the corresponding coherent structures are presented in~\autoref{fig:expLoops}. The two main recirculating vortices are identified. The set of particles orbiting around the outside of the domain is also identified.  For the experiment, particles that pass through the chaotic sea only during the final periods also belong to this outer structure because not enough mixing with the structure has occurred for some of these particles to be excluded from the orbit structure. For all three cases, the braided length of the topological loop \(\abs{B\ell}\)  either shrinks or remains the same.  The left and right recirculating vortex structures correspond to ratios of 1 and 0.86, respectively, while the orbiting structure ratios 0.92, demonstrating that perceivably-coherent structures result in roughly-constant lengths of topological loops. Having demonstrated the coherent structures approach to the PT trajectories, the analysis was reiterated on datasets of synthetic PIV trajectories. Each recirculating vortex was detected with resolution dependent on the trajectory initialization. The Supplemental Material Figure \citep{SuppMat} presents the coherent structures detected from two sample sets of synthetic trajectories, one with 33 trajectories and one with 100 trajectories.

\subsection{Comparison}\label{sec:comparison}

These results demonstrate the ability of the braid theory approach to detected coherent structures accurately and robustly in an experimental, periodically forced flow. The method found similar qualitative results using the experimental and analytic model trajectories. Despite the lower particle density in the experimental system, the coherent structures representing the vortex edges were detected, even with the structures containing as few as three particles, which shows the reliability of the method even for sparse datasets.  The small-scale features detected in the model flow were not identified from the experimental data because too few particles were seeded in those regions.

When comparing the braid-based coherent structures to the Poincar\'e map for the model flow there are similarities and differences.  First, the main recirculating vortices, represented in red and blue for both the model and experimental examples, correspond to the islands identified in \autoref{fig:poincare-map}(b). The second feature identified is an outer structure corresponding to a set of particles that are advected at very low velocities around the main vortex system. This outer structure encompassed a much higher proportion of particles for the experimental flow than for the theoretical flow. This may be due to the sparsity of the dataset: if fewer trajectories are present, then they may be less likely to entangle with the chaotic sea.

To improve the resolution of the braid-based coherent structures, more particle trajectories are needed to improve the spatiotemporal coverage. In laboratory experiments, however, there are two main limitations to accomplishing this. Due to imagery and tracking issues, the trajectories of several experimental particles could not be fully reconstructed for all periods. This is why only 33 out of 80 trajectories were retained for the analysis.  The braid approach requires all trajectories to span the same time interval and it needs a continuous dataset, i.e., no missing datapoints within trajectories. The imagery issue was mostly due to the challenge of having a contrast between particles and background flow that is high enough for particles to be detected at each time step. The tracking issue was mostly due to the wide range of velocities between particles located close to the rotor and particles in the outer region, which made the construction of trajectories from their respective particles difficult. The other main challenge to obtaining a sufficient spatiotemporal coverage is that as the particle density increases the particle-particle interactions become more important and can cause the flow to significantly deviate from the analytic model. Some of the features of interest present in the flow being small-sized, it was therefore difficult to seed the spatial domain to discern all of them. Nonetheless, the remaining trajectories were already sufficient to succeed at detecting the largest coherent structures using the braid theory approach.

\section{Estimating the growth rate of material interfaces}\label{sec:results-mixing}

Newhouse and Yomdin~\cite{Newhouse1993,Newhouse1988,Yomdin1987} established that, in planar flows, the exponential growth rate of material interfaces is equal to the topological entropy of the flow. In turn, topological entropy of the flow is bounded below by braid entropy, a measure of complexity of braids of any periodic trajectories in the flow~\cite{Boyland2000,Gouillart2006}. Adding trajectories to a braid can only increase the braid entropy, therefore a reasonable approximation strategy involves computing braid entropies of braids comprising more and more trajectories in an effort to narrow the gap between the braid and topological entropy. Applying this strategy to experimentally-recorded or simulated trajectories is difficult to do directly, as most trajectories are not periodic. Nevertheless, \citet{Budisic2015} demonstrated that computing a so-called Finite-Time Braiding Exponent (FTBE) provides a reasonable data-driven estimate of braid entropy and consequently an estimate of the rate of exponential growth of material interfaces. In this section, we apply this procedure to evaluate the degree of correspondence between the experimental and model rotor-oscillator flows.

\subsection{Theory}
\label{sec:alg-ftbe}

The most straightforward way to quantify the complexity of a braid is to count the number of crossings in the braid, while avoiding counting trivial crossings. For example, two braids with two strands each:
\begin{center}
\begin{tikzpicture}
  \braid[rotate=-90,
  width=10pt, % spacing between strands
  height=10pt, % size of element of the braid (crossing)
  style strands={1}{blue!30!black,line width=2pt},
  style strands={2}{blue!30!black,black},
  style strands={3}{green!30!black,line width=2pt},
  style strands={4}{green!30!black,black}]
  s_1-s_3 s_1-s_3 s_1 s_1^{-1}  s_1-s_3 s_1-s_3;
\end{tikzpicture}
\end{center}
both have 3 crossings, since the two central crossings in the top braid can be disentangled even with the endpoints pinned. More rigorously, before counting crossings the braid is reduced by employing braid group rules that cancel out such intersections. Define \emph{braid length} \(\# B\) to be the number of intersections after the braid is reduced.  Given a topological braid \(B_{N, T}\) that corresponds to \(N\) trajectories collected over a time interval of duration \(T\), the average \emph{braid growth rate} of \(B_{N,T}\) is
\begin{equation}
  \label{eq:growth}
  G(B_{N,T}) = \frac{\# B_{N,T}}{T}.
\end{equation}

The number of intersections alone is a poor indicator of complexity.  For example, the two 3-stranded braids
\begin{center}
  \begin{tikzpicture}
    \braid[rotate=90,
    xscale=-1, % flip l-r
  width=10pt, % spacing between strands
  height=10pt, % size of element of the braid (crossing)
  style strands={thick},
  line width=1pt,number of strands=6,
  style strands={1}{blue!30!black},
  style strands={2}{blue!30!black},
  style strands={3}{blue!30!black,line width=2pt},
  style strands={4}{green!30!black},
  style strands={5}{green!30!black},
  style strands={6}{green!30!black,line width=2pt}]
  s_1-s_4 s_2-s_5^{-1}
  s_2-s_4 s_2-s_5^{-1}
  s_2-s_4 s_2-s_5^{-1}
  s_2-s_4 s_2-s_5^{-1};
\end{tikzpicture}
\end{center}
have the same length, but the strands in the bottom braid are intertwined in a more intricate manner.

The \emph{topological entropy of a braid}, or simply \emph{braid entropy}, measures the complexity of a braid by quantifying how fast loops grow as they are slid along the braid (as in \autoref{fig:sample-loops}). Braid entropy can be computed by finding the largest rate of growth of a topological loop under a repeated action of the braid~\cite{Thiffeault2005,Moussafir2006}. When the physical braid comprises periodic trajectories, the entropy of the associated topological braid is a lower bound for the topological entropy of the flow. This property was successfully  exploited to create stirring protocols for rod-based stirrers~\cite{Boyland2000, Thiffeault2006, Finn2011, Thiffeault2018}, to design spatially periodic mixers~\cite{Finn2006}, and to analyze stirring of a passive tracer by compact coherent structures (so-called \emph{ghost rods})~\cite{Gouillart2006,Thiffeault2006, Grover2012,Stremler2011}.

When trajectories in the physical braid are not periodic, the theory is less developed compared to the periodic case, and published works rely on numerical simulations~\cite{Thiffeault2005,Thiffeault2010}. Nevertheless, when trajectory segments are long enough, it is possible to define the Finite-Time Braiding Exponent (FTBE)~\cite{Budisic2015}, which measures the exponential growth rate of a particular topological loop during a single application of a braid. (By comparison, topological entropy measures asymptotic exponential growth with respect to repeated application of the braid.)

\begin{figure}[h!]
  \centering
    \includegraphics[width=0.75\linewidth]{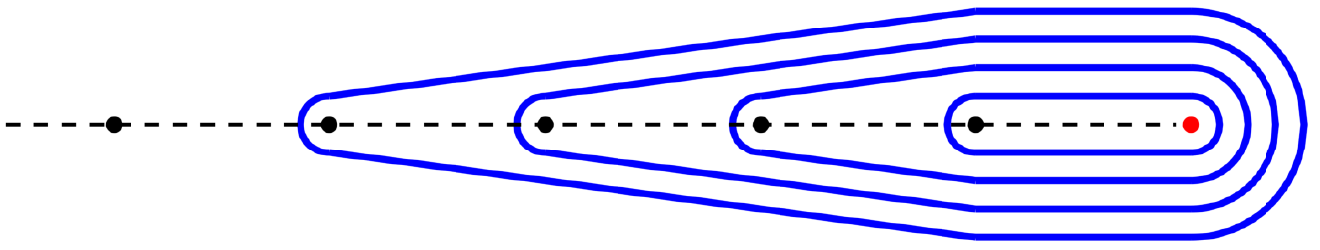}

\vspace{12pt}

    \includegraphics[width=0.75\linewidth]{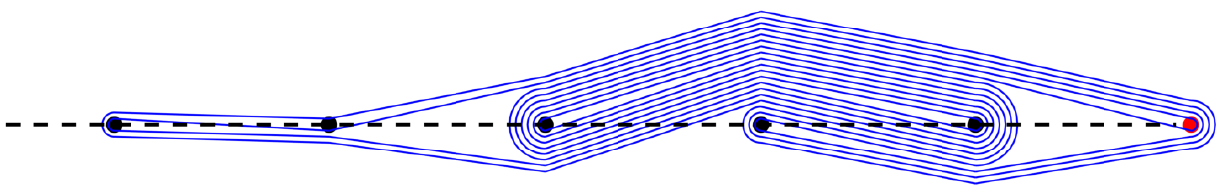}
  \caption[]{Top: Loop~$\ell_0$ used to compute the Finite-Time Braiding
    Exponent, with Dynnikov vector \(\ell_{0}= \begin{bsmallmatrix}
    0 & 0 & 0 & 0 & -1 & -1 & -1 & -1
  \end{bsmallmatrix}\).
  Bottom: The loop after action of the braid from \autoref{fig:sample-braid}, with Dynnikov vector \( B\ell_{0}= \begin{bsmallmatrix}  -1 &0 &8 &1 &0 &-6 &-3 &7  \end{bsmallmatrix}\).  Black dots: strands belonging to the braid; right-most red dot: the
    auxiliary ``anchor'' strand. The length of a loop is the number of intersections with the dashed axis (\(\abs{\ell_0} = 4, \abs{B\ell_{0}}=36\)).}
  \label{fig:dw-action}
\end{figure}

The Finite-Time Braiding Exponent is given by
\begin{equation}
  \label{eq:ftbe}
  \ftbe(B_{N,T}) = \frac{1}{T}\log\frac{\abs{B_{N,T}\ell_0}}{\abs{\ell_0}},
\end{equation}
where \(\ell_0\) is the topological loop in \autoref{fig:dw-action} (top). The particular loop \(\ell_{0}\) is chosen so that any nontrivial braid results in a nonzero deformation~\cite{Budisic2015}. To ensure this, the components of the loop are ``anchored'' on the boundary of the domain, which is topologically equivalent to adding an additional strand  (depicted as the hollow circle) that does not participate in dynamics. Justifications for the choice of \(\ell_{0}\) is given at length in \cite{Dynnikov2007,Dehornoy2008}.

Compared to \eqref{eq:pair-loop-growth}  in section \ref{sec:results-lcs}, the FTBE is calculated for the specific loop \(\ell_0\) instead of a set of simple pair-loops.   Further information on FTBE calculations can be found in~\cite{Budisic2015}.

Compared to the connection between topological entropy and the rate of material growth, the connection between FTBEs and the rate of material growth is more heuristic. Nevertheless, numerical investigations in a chaotic flow~\cite{Budisic2015} shows that FTBEs can indeed be used as a reasonable proxy to the topological entropy of a braid when in the absence of periodic Lagrangian trajectories from the flow. Consequently, FTBEs characterizes the rate of growth of material lines in this work. Braid growth \(G\) will be used as the independent variable in comparing FTBE of braids with different number of strands.

\subsection{Comparison of the model and experimental flows}\label{sec:FTBE-num}

We compare the complexity of numerical trajectories in the model and experimental velocity fields. The parameters of the PIV experiment in \autoref{tab:exp-parameters} correspond to a nondimensional oscillation period $\tau = 5$  and non-dimensional parameters $\epsilon = 0.1250$ and $\lambda = 1.2566$. We initialized trajectories in the chaotic region where primary and secondary vortices meet (see \autoref{fig:streamlines}), as this is where exponential growth of material lines is expected, and simulated them for 30 oscillation periods (\(T = 30\tau = 150\)). As the PIV velocity field was recorded over 11 oscillation periods (see \ref{sec:roflow-exp-piv}), to achieve \(T = 150\) we repeated three times the first 10 oscillation periods.

\begin{figure}[ht]
  % \graphicspath{ {./images/} }
  \subfigure[\ Samples of analytic model velocity field trajectories]{
  \begin{tikzpicture}
    \begin{axis}[y dir=reverse,
      axis on top,
      % xticklabel pos=top,
      xtick={-3,-2,...,3},
      xmin=-3,xmax=3,
      ymin=-1,ymax=1,
      axis equal image,
      grid=none, %both
      grid style={green!75!black,dashed},
      width=\linewidth,
      ylabel style={rotate=-90},
      xlabel={\(y\) },
      ylabel={\(x\) }
      ]
      \addplot[plot graphics/node/.append style={yscale=-1,anchor=north west}] graphics [xmin=-3,xmax=3,ymin=-1,ymax=1] {\ftbedataset\detokenize{NUM_poincare}.png};
      \draw[line width=1.5pt,color=red!90!black] (-0.5,-0.25) rectangle (0.5,-0.5);
    \end{axis}
  \end{tikzpicture}}

  \subfigure[\ Samples of PIV velocity field trajectories]{
  \begin{tikzpicture}
    \begin{axis}[y dir=reverse,
      axis on top,
      % xticklabel pos=top,
      xtick={-3,-2,...,3},
      xmin=-3,xmax=3,
      ymin=-1,ymax=1,
      axis equal image,
      grid=none, %both
      grid style={green!75!black,dashed},
      width=\linewidth,
      ylabel style={rotate=-90},
      xlabel={\(y\) },
      ylabel={\(x\) }
      ]
      \addplot[plot graphics/node/.append style={yscale=-1,anchor=north west}] graphics [xmin=-3,xmax=3,ymin=-1,ymax=1] {\ftbedataset\detokenize{PIV_poincare}.png};
      \draw[line width=1.5pt,color=red!90!black] (-0.5,-0.1) rectangle (0.5,-0.5);
    \end{axis}
  \end{tikzpicture}}
\caption{Poincar\'{e} plots of trajectories in braids used to compare FTBE and braid growth rate between analytic model and PIV velocity fields. Trajectories were initialized uniformly inside the highlighted rectangles
  (a) \((x,y) \in [-0.25,-0.5] \times [-0.5,0.5]\) for model velocities;
  (b) \((x,y) \in [-0.1,-0.5] \times [-0.5,0.5]\) for PIV velocities.
The Poincar\'{e} map is taken at the rotor's oscillation period \(\tau\).}
	\label{fig:ftbe-poincare}
\end{figure}
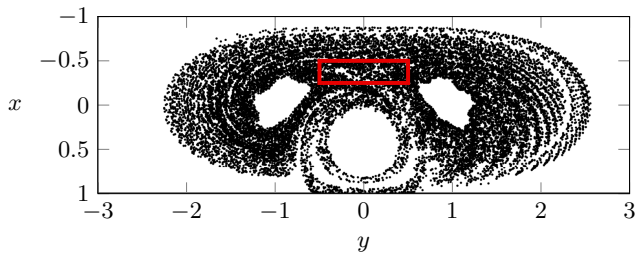
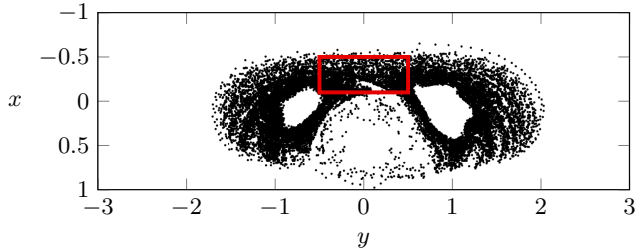

Integrating PIV velocity field for long times compounds measurement errors in the velocities, especially in the slow regions of the flow, resulting in unphysical trajectories, e.g., trajectories that exit the domain. We removed such trajectories from the PIV dataset and replaced them with additional  simulated trajectories until we obtained a full set of \(N=600\) well-behaved trajectories. We collected the same number of trajectories from the model velocity field. Even though the model and PIV velocity fields are qualitatively similar (see \autoref{fig:streamlines}), the precise location of the saddle-like point sitting between the primary and secondary vortices is sligthly different. To ensure that trajectories are seeded in the chaotic zone in both cases, we used a different initialization rectangle for the two velocity fields. Consequently, Poincar\'{e} plots in \autoref{fig:ftbe-poincare} show that both datasets have similar qualitative features in the region of interest.   Finally, we converted sets of trajectories into braids using the MATLAB toolbox \texttt{braidlab}~\cite{braidlab}.

The braids of trajectories are characterized by their braid growth rates~\eqref{eq:growth} and FTBE values~\eqref{eq:ftbe}, which respectively quantify how many non-trivial strand crossings are generated, and the amount of stretching that these crossings impose on topological material lines. We analyze ensembles of \(S=500\) braids, each with \(n\) strands, in order to estimate distributions of FTBE and braid growth. By randomly choosing \(n\) trajectories out of the full set of \(N=600\) simulated trajectories, for each number of strands \(n=3,\dots, N/2=300\), we create \(S\) realizations of braids \(B_{n,T}\). By reusing trajectories in several braids, fewer trajectories need to be simulated, avoiding the costly initialization and resimulation of trajectories for individual realizations. Further discussion of the resampling step is in \cite{Budisic2015}.

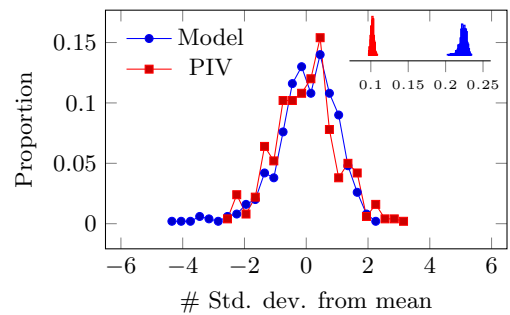
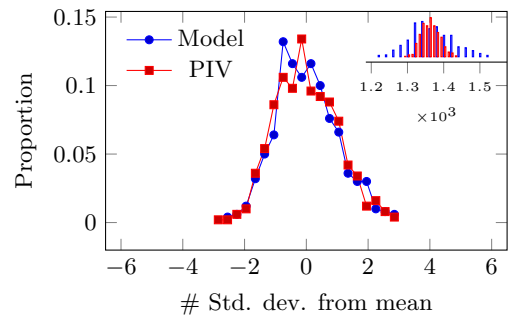
\begin{figure}[h!]
  \centering
  \subfigure[\ FTBE]{
    \begin{tikzpicture}
      \begin{axis}[width=0.8\linewidth, height=0.55\linewidth,
        enlargelimits=0.15,
        xmin=-5,xmax=5, mark size=1.5pt,
        legend pos = north west, legend style={fill=none,draw=none},
        yticklabel style={/pgf/number format/.cd,fixed,precision=2},
        xlabel={\# Std. dev. from mean},
        ylabel={Proportion}]
        \addplot+ table[
        x=Value,
        y=Proportion,
        col sep=comma]{FTBEnum_histStd.csv};
        \addlegendentry{Model};
        \addplot+ table[
        x=Value,
        y=Proportion,
        col sep=comma]{FTBEpiv_histStd.csv};
        \addlegendentry{PIV};
      \end{axis}

      \begin{axis}[xshift=0.375\linewidth,yshift=2.5cm,
        y axis line style={draw=none},
        width=0.4\linewidth,height=2.25cm,ybar,
        bar width=0.5pt,
        axis x line*=bottom,
        x tick label style={font=\tiny},
        ymajorticks=false,
        enlargelimits=0.15]
        \addplot+ table[
        x=Value,
        y=Proportion,
        col sep=comma]{FTBEnum_histRaw.csv};
        \addplot+ table[
        x=Value,
        y=Proportion,
        col sep=comma]{FTBEpiv_histRaw.csv};
      \end{axis}

    \end{tikzpicture}
    }

  \subfigure[\ Growth rate]{
    \begin{tikzpicture}
      \begin{axis}[width=0.8\linewidth, height=0.55\linewidth,
        xmin=-5,xmax=5, mark size=1.5pt,
        enlargelimits=0.15, ymin=0,
        xmin=-5,xmax=5,
        ylabel={Proportion},
        yticklabel style={/pgf/number format/.cd,fixed,precision=2},
        legend pos = north west,legend style={fill=none,draw=none},
        xlabel={\# Std. dev. from mean }]
        \addplot+ table[
        x=Value,
        y=Proportion,
        col sep=comma]{CLRnum_histStd.csv};
        \addlegendentry{Model};
        \addplot+ table[
        x=Value,
        y=Proportion,
        col sep=comma]{CLRpiv_histStd.csv};
        \addlegendentry{PIV};
      \end{axis}
      \begin{axis}[xshift=0.4\linewidth,yshift=2.5cm,
        y axis line style={draw=none},
        width=0.4\linewidth,height=2.25cm,
        ymajorticks=false,
        axis x line*=bottom,
        x tick label style={font=\tiny},
        xlabel={\tiny $\times 10^{3}$},
        xticklabel={$\pgfmathprintnumber
          [sci,sci generic={mantissa sep=,exponent={}}]
          {\tick}$
        },
        ybar,
        bar width=0.5pt,
        enlargelimits=0.15]
        \addplot+ table[
        x=Value,
        y=Proportion,
        col sep=comma]{CLRnum_histRaw.csv};
        \addplot+ table[
        x=Value,
        y=Proportion,
        col sep=comma]{CLRpiv_histRaw.csv};
      \end{axis}

    \end{tikzpicture}
  }
  \caption{Standardized distributions of FTBEs and growth rates with \(S=500\) samples of \(n=300\)-stranded sub-braids (out of \(N=600\) trajectories, \(T=30\tau=150\)); non-standardized histograms are inset into each graph. \label{fig:distributions}}
\end{figure}

\autoref{fig:distributions} shows histograms of FTBEs and braid growth rates for both sets of braids with $S=500$ realizations of the largest number of strands \(n=300\), subsampled from the full simulated set of \(N\). From a qualitative comparison with Poincar\'{e} plots, one would expect similar results for both model and PIV velocity fields. The shapes of distributions are similar, but with quantitative differences. The model velocity field shows a greater variance in the growth rate, while FTBE distributions are non-overlapping, indicating a significant separation of recorded FTBE values. The model velocity field may allow initialization of rare-event trajectories more easily, e.g., in intricate periodic islands that would be washed out in experimental system, potentially explaining the variance in growth. The distance between FTBE distributions appears to be more significant, indicating that even though the number of strand intersections is comparable, the intersections for the model flow generate noticeably more complexity than the experiment.

Presently, we do not know of a trajectory mechanism that would explain this difference. We do not think that the difference in initialization window (see \autoref{fig:ftbe-poincare}) could cause this deviation, as trajectories in both cases occupy the chaotic region surrounding the vortices despite the slight numerical differences in locations of those regions.

Additionally, artificially degrading the quality of the model does not change the outcome. We have sampled the model velocity field on the same grid as the PIV velocity field, and performed the same space-time interpolation that was used for the PIV velocity field. The resulting interpolated model velocity field had the same FTBE values as the continuous model velocity field, suggesting that discretization and interpolation themselves are not the cause for the lower FTBE values in the PIV flow. These results are not presented in figures, as they do not differ from the model flow results.

Ultimately, the quantitative difference between FTBE values may speak to the sensitivity of the braid dynamics techniques, in particular FTBE, to differences between flows that are otherwise challenging to determine.

In order to exhibit the dependence of distributions of FTBEs and growth rates on duration \(T\) and number of strands \(n\) for braids  \(B_{n,T}\), we compute summary statistics and graph them against \(T\) and \(n\). FTBE distributions are represented by their maximum values, as braid entropy of \emph{any} braid is a lower bound for the topological entropy of the flow. Growth rate distributions are simply represented by mean values, as there are no special considerations for this quantity.

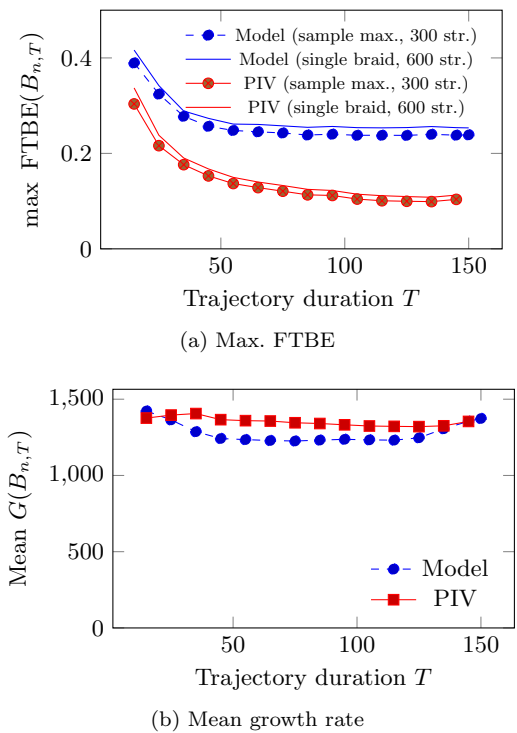
\begin{figure}[htb]
  \centering
\subfigure[ \ Max.~FTBE]{
    \begin{tikzpicture}
      \begin{axis}[width=0.8\linewidth, height=0.55\linewidth,
        xlabel={Trajectory duration \(T\)},
        ylabel={\(\mftbe[90](B_{n,T})\)}, ymin=0, ymax=0.5,
        legend pos = north east,legend style={fill=none, draw=none, nodes={scale=0.75, transform shape}}]
        \addplot+[dashed] table[
        x=Ts,
        y=Max,
        col sep=comma]{FTBEnum_vsTime.csv};
        \addlegendentry{Model (sample max., $300$ str.)};
        \addplot+[mark=none,color=blue] table[
        x=Ts,
        y=FTBE,
        col sep=comma]{NUMFull.csv};
        \addlegendentry{Model (single braid, $600$ str.)};
        \addplot+[color=red] table[
        x=Ts,
        y=Max,
        col sep=comma]{FTBEpiv_vsTime.csv};
        \addlegendentry{PIV (sample max., $300$ str.)};
        \addplot+[mark=none,color=red] table[
        x=Ts,
        y=FTBE,
        col sep=comma]{PIVFull.csv};
        \addlegendentry{PIV (single braid, $600$ str.)};

      \end{axis}
    \end{tikzpicture}
  }
  \subfigure[\ Mean growth rate]{
    \begin{tikzpicture}
      \begin{axis}[width=0.8\linewidth, height=0.55\linewidth,
        xlabel={Trajectory duration \(T\)},
        ylabel={Mean \(G(B_{n,T})\)}, ymin=0,
        legend pos = south east,
        legend style={fill=none,draw=none}]
        \addplot+[dashed] table[
        x=Ts,
        y=Mean,
        col sep=comma]{CLRnum_vsTime.csv};
        \addlegendentry{Model};
        \addplot+ table[
        x=Ts,
        y=Mean,
        col sep=comma]{CLRpiv_vsTime.csv};
        \addlegendentry{PIV};
      \end{axis}
    \end{tikzpicture}
  }
  \caption{Dependence of the maximum FTBE and mean growth rate on trajectory duration \(T\). The comparison is between the maximum FTBE over a sample of \(n=300\)-stranded braids and a single braid of \(n=600\) strands, all from analytic model and PIV flows.\label{fig:evol-in-time}}
\end{figure}

The maximum FTBE in a sample should provide the best-available bound to topological entropy for sufficiently-long duration \(T\)~\cite{Budisic2015}. \autoref{fig:evol-in-time} demonstrates that the values of maximum FTBE and mean growth stabilize well ahead the end of simulations, which leads us to conclude that studying trajectories of length \(T = 30\tau\) is representative of values in the long-time limit.

\begin{figure}[htb]
  \centering
  \subfigure[\ FTBE of sub-braids (\(T = 150\)) \label{fig:FTBE-vs-n}]{
    \begin{tikzpicture}
      \begin{axis}[width=0.8\linewidth, height=0.55\linewidth,
        xlabel=Number of strands \(n\),
        ylabel=\(\mftbe(B_{N,T})\), ymin=0,
        legend pos = south east,legend style={fill=none,draw=none}]
        \addplot+[dashed] table[
        x=Ns,
        y=Max,
        col sep=comma]{FTBEnum_vsStrands.csv};
        \addlegendentry{Model};
        \addplot+ table[
        x=Ns,
        y=Max,
        col sep=comma]{FTBEpiv_vsStrands.csv};
        \addlegendentry{PIV};
        \addplot[mark=o,color=blue] coordinates {(600,0.253388014699152)}; % from NUMFull.csv
        \addplot[mark=square,color=red] coordinates {(600,0.112650479439487)}; % from PIVFull.csv
      \end{axis}
    \end{tikzpicture}
  }
  \subfigure[\ Growth rate of sub-braids (\(T = 150\) ) \label{fig:growth-vs-n}]{
    \begin{tikzpicture}
      \begin{axis}[width=0.8\linewidth, height=0.55\linewidth,
        xlabel={Number of strands \(n\)}, xmode=log, ymin=0,
        ylabel={\(G_{\overline{n}}(B_{N,T})\) }, ymode=log,
        legend pos = north west,legend style={fill=none,draw=none}]
        \addplot+[dashed] table[
        x=Ns,
        y=Max,
        col sep=comma]{CLRnum_vsStrands.csv};
        \addlegendentry{Model};
        \addplot+ table[
        x=Ns,
        y=Max,
        col sep=comma]{CLRpiv_vsStrands.csv};
        \addlegendentry{PIV};
        \logLogSlopeTriangle{0.9}{0.2}{0.55}{2}{black!75!white}; % slope triangle at (0.9, 0.55), width=0.2, slope=2
        \addplot[mark=o,color=blue] coordinates {(600,5514.62)}; % from NUMFull.csv
        \addplot[mark=square,color=red] coordinates {(600,5430.73793103448)}; % from PIVFull.csv
      \end{axis}
    \end{tikzpicture}
  }

  \caption{Effect of increase in braid size (number of strands) on FTBE and growth rate of braids of trajectories from model and PIV flows.\label{fig:evol-in-strands} Filled-in marks correspond to maximum of a sample of \(n\)-stranded braids. Hollow marks correspond to the single \(600\)-stranded braid.}
\end{figure}
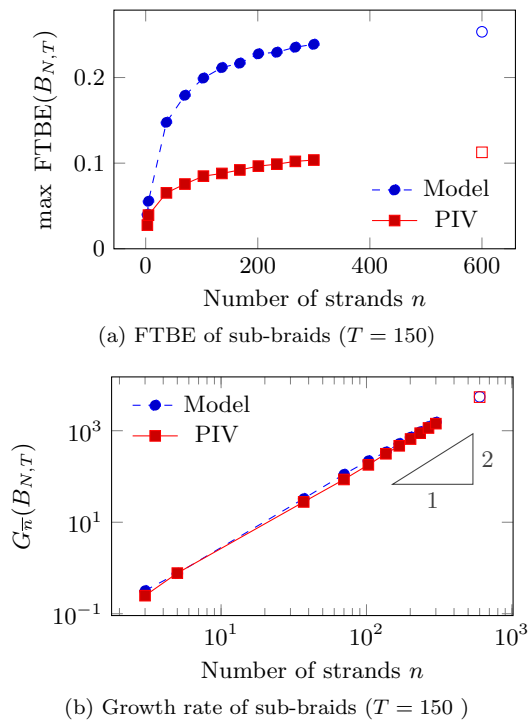

Intuitively, the more strands participate in a braid, the higher the potential
for complexity. Consequently in chaotic regions, where we initialized trajectories, both FTBE and braid growth should increase with \(n\). \autoref{fig:evol-in-strands} shows the observed dependence. The growth rate for both flows increases quadratically with the number of strands, which can be explained by noting that \(n\) strands can form \(\binom{n}{2}=\mathcal{O}(n^{2})\) pairs. In chaotic regions, statistical quantities tend to mimic those in stochastic processes with independent increments, so over large periods of time any of those \(\binom{n}{2}\) possible crossings should be equally likely. As more pairs are confined in a finite space, we observe a quadratic increase in the number of crossings for a fixed time interval.

FTBE increases with \(n\) in both velocity fields, following the shape previously observed in~\cite{Budisic2015} for which there is no good mathematical model, as far as we are aware. Topological entropy of the flow acts as the upper bound for the curve, but it is not clear if the bound is tight. In planar flows, topological entropy corresponds to the maximal rate of exponential growth of material lines. Since the FTBE is non-zero, we infer that both flows exhibit exponential growth of material interfaces, although the FTBE estimates of these rates differ quantitatively.  We do not know if the difference is dynamically significant.  Future work could contrast FTBE results to those obtained by the recently-developed eTEC technique~\cite{Roberts2019} as an independent comparison.

Finally, we point out some challenges encountered here. An effective comparison of complexity based on braid dynamics requires Lagrangian trajectories that cover many periods of oscillations, which is experimentally challenging for several reasons. As explained in Section~\ref{sec:roflow-exp-piv}, the analytical model assumes that the rotor is a point vortex and that the Reynolds number is zero. Additionally, for the FTBE to be computed, the motors of the rotor and the oscillator must run for many hours, which inevitably introduces some amount of thermal energy into the flow and slightly changes its viscosity over time. For these reasons, we only used 10 periods of experimental data to reconstruct the flow, which required artificially repeating the velocity field to obtain Lagrangian trajectories over 30 periods. The chosen periods were the first 10 \emph{full} periods after the first peak of oscillation, discarding the transient dynamics after the initial acceleration.
%Finally, we point out some challenges encountered here. An effective comparison of complexity based on braid dynamics requires Lagrangian trajectories that cover many periods of oscillations, which is experimentally challenging for two main reasons. First, the motors of the rotor and the oscillator must run for many hours, which inevitably introduces some amount of thermal energy into the flow and slightly changes its viscosity over time. Second, the wide range of flow velocities between the area close to the rotor and in the outer region limits the PIV analyses (Section~\ref{sec:roflow-exp-piv}).  For these reasons, we only used 10 periods of experimental data to reconstruct the flow, which required artificially repeating the velocity field to obtain Lagrangian trajectories over 30 periods.

\section{Discussion}
\label{sec:discussion}

We demonstrated that methods based on braid dynamics can identify qualitative
features of material transport in model and experimental rotor-oscillator
flows. In particular, Section~\ref{sec:results-lcs} shows that braid dynamics
can distinguish between nearby transport structures---secondary vortices and a
recirculating region---even when only a few Lagrangian trajectories are seeded in them, with structures being detected from as few as two trajectories.  For quantities measuring material deformation (FTBE and growth rate), we demonstrated in Section~\ref{sec:results-mixing} that our two qualitatively-similar flows result in similar distributions, with parallel trends in their time-evolution.

Based on these results, we expect that the braid dynamics methods will be useful in situations where only a small number of Lagrangian trajectories from a two-dimensional flow is available, either due to experimental challenges, or because only a rough estimate of flow properties is desired.

During the comparison of model and experimental data, it became clear that processed data needs to satisfy, at a minimum, the following requirements in order to be successfully processed using computational braid dynamics methods:
\begin{compactitem}
  \item There cannot be significant gaps in the discrete time series of particle tracks;
  \item At least a few particles should be seeded in each structure to be detected;
  \item Quantitative comparison of FTBE values is unreliable in regions without significant mixing;
  \item Reliable calculation of FTBEs requires either long Lagrangian trajectories, or measured velocity fields that can be used to generate such trajectories.
  \end{compactitem}

  As explained in Sections~\ref{sec:comparison} and~\ref{sec:FTBE-num}, we have had to overcome several difficulties in order to obtain a set of experimental particle tracks that were both sufficiently dense and gapless.  Other experimental setups may face similar challenges, which currently limits the range of application of methods based on braid dynamics. Nonetheless, in experimental conditions that do satisfy the requirements, braid dynamics techniques can be competitive with alternative techniques for identifying coherent structures or quantifying the amount of mixing in a two-dimensional flow. We do point out that the recently-developed eTEC technique~\cite{Roberts2019} shows promise in overcoming at least some of the obstacles mentioned here.

  We highlight three directions for future work to increase the usefulness of braid dynamics techniques for experimental data. First, many more existing (gappy) datasets could be processed if there was a technique for filling-in missing segments in Lagrangian trajectories that do not introduce significant additional entropy. Second, fewer strands would be needed if there existed a theoretical model for growth of FTBEs with number of strands. Finally, the efficiency of search for coherent structures could be improved by considering the geometry of the space of topological loops.  Some of these directions may be amenable to modern machine learning techniques.

\begin{acknowledgments}
We thank Gustaaf (Guus) Jacobs for his help running CFD models for preliminary analysis of the experimental setup, Nicholas Ouellette and Douglas Kelley for their assistance with processing PIV data and Greg Voth for a helpful discussion about the comparison between the PIV data and the model. This research was supported by the U.S.~National Science Foundation (NSF), under Grants No. CMMI-1233935 and AGS-1520825.
\end{acknowledgments}

\bibliography{braids_APS}

\end{document}

\title{Supplementary material to\\ ``Using braids to quantify interface growth and coherence in a rotor-oscillator flow''}

%\thanks{A footnote to the article title}%

\author{Margaux Filippi}
\email{margaux@mit.edu}

\author{Marko Budi\v{s}i\'{c}}%

\author{Michael R. Allshouse}
\author{S\'{e}verine Atis}

\author{Jean-Luc Thiffeault}

\author{Thomas Peacock}

\date{\today}% It is always \today, today,
             %  but any date may be explicitly specified

\maketitle

The particle tracking experiment directly provided the set of trajectories analyzed to find coherent structures. It is also possible to detect coherent structures from Particle Image Velocimetry (PIV) by generating a synthetic set of trajectories from the measured velocity field.

\begin{figure}[H]
	\centering
	\includegraphics[width=\linewidth]{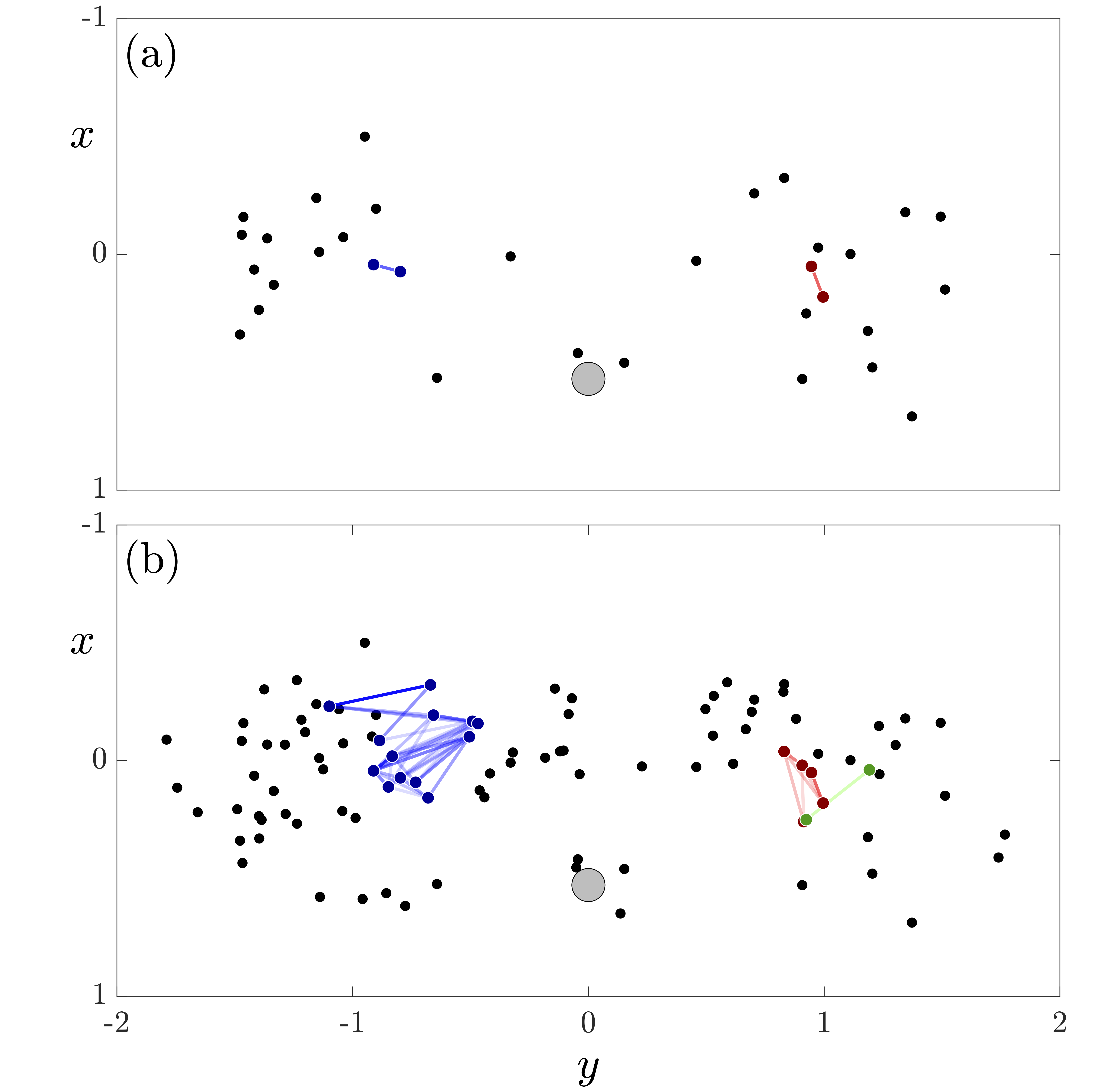}
	\caption{Coherent structures analysis on datasets of synthetic PIV trajectories. Each recirculating vortex was detected with resolution dependent on the trajectory initialization. (a) The analysis of 33 trajectories detected a blue coherent structure within the left recirculating vortex and a red coherent structure within the right recirculating vortex, each structure containing 2 trajectories. (b) The analysis of 100 trajectories detected a blue coherent structure on the left with 13 trajectories, a red coherent structure on the right with 5 trajectories and a green coherent structure within the right vortex with 2 trajectories.}
	\label{fig:supfig}
\end{figure}

Using the same 10 periods of data used to create the synthetic trajectories for the FTBE analysis, two sets of 33 and 100 trajectories are generated. To prepare the initial conidtions, particles are uniformly seeded in $-1.5 < x < 1.5$, $-0.5 < y < 0.2$, then advected for 5 initial periods to minimize the influence of the initial seeding. After such initialization, the trajectories are recorded for 50 periods of wall-oscillation and converted to braids, which are then used in detection of coherent structures as explained in Sec.~IV of the paper. The resulting coherent structures and the network of non-growing pair loops are presented in \autoref{fig:supfig}.

The smaller data set produces a similar result to the one found using the particle tracking trajectories where each vortex contains only two particles. The larger dataset include many particles in each of the recirculating vortex. The perceived increase in the vortex footprint demonstrates the importance of highly sampling the coherent structure to get a better approximation for the feature.

\vfill\phantom{ } % avoid floating last paragraph to the bottom of the column

